\documentclass[conference]{IEEEtran}
\IEEEoverridecommandlockouts
\usepackage{cite}
\usepackage{amsmath,amssymb,amsfonts}
\usepackage{textcomp}
\usepackage{xcolor}
\usepackage[pdftex]{graphicx}
\usepackage{subcaption}
\usepackage{braket}
\usepackage{algorithm}
\usepackage{algpseudocode}
\usepackage{algorithmicx}
\usepackage[scientific-notation=true]{siunitx}
\usepackage{url}
\usepackage{fancyhdr}
\usepackage[bookmarks=false]{hyperref}
\usepackage{enumerate}
\usepackage{multirow}
\usepackage{color}
\definecolor{backcolour}{rgb}{0.97,0.97,0.97}
\definecolor{myblue}{RGB}{71,124,202}
\definecolor{commentcolour}{rgb}{0.2,0.2,0.6}
\definecolor{codegray}{rgb}{0.5,0.5,0.5}
\definecolor{codepurple}{rgb}{0.58,0,0.82}
\usepackage{listings}

\lstdefinelanguage{Scaffold}{%
  language     = C,
  morekeywords = {module, qbit},
}
\lstdefinestyle{mystyle}{
    backgroundcolor=\color{backcolour},   
    commentstyle=\color{magenta},
    keywordstyle={\bfseries\color{myblue}},
    numberstyle=\tiny\color{codegray},
    stringstyle=\color{codepurple},
    basicstyle=\ttfamily\footnotesize,
    breakatwhitespace=false,         
    breaklines=true,
    mathescape=true,
    captionpos=b,                    
    keepspaces=true,                 
    numbers=left,                    
    numbersep=5pt,                  
    showspaces=false,                
    showstringspaces=false,
    showtabs=false,                  
    tabsize=2
}
\def\BibTeX{{\rm B\kern-.05em{\sc i\kern-.025em b}\kern-.08em
    T\kern-.1667em\lower.7ex\hbox{E}\kern-.125emX}}
\begin{document}

\title{{\fontsize{21.5}{25.8}\selectfont Magic-State Functional Units: Mapping and \\Scheduling Multi-Level Distillation Circuits for\\\vspace{-0.1cm} Fault-Tolerant Quantum Architectures}
}

\author{
Yongshan Ding\IEEEauthorrefmark{1}\IEEEauthorrefmark{4}\thanks{\IEEEauthorrefmark{4}These two authors contributed equally.},
Adam Holmes\IEEEauthorrefmark{1}\IEEEauthorrefmark{4},
Ali Javadi-Abhari\IEEEauthorrefmark{2},\\
Diana Franklin\IEEEauthorrefmark{1}, 
Margaret Martonosi\IEEEauthorrefmark{3} and
Frederic T. Chong\IEEEauthorrefmark{1}
\\[1em]
\begin{tabular}{*{3}{>{\centering}p{.31\textwidth}}}
\IEEEauthorrefmark{1}\textit{University of Chicago}& \IEEEauthorrefmark{2}\textit{IBM T.J. Watson Research Center}& \IEEEauthorrefmark{3}\textit{Princeton University}\tabularnewline
\{yongshan, adholmes, dmfranklin, chong\}@uchicago.edu & ali.javadi@ibm.com & mrm@princeton.edu 
\end{tabular}}

\maketitle

\begin{abstract}
Quantum computers have recently made great strides and 
are on a long-term path towards useful fault-tolerant computation.
A dominant overhead in fault-tolerant quantum computation is the
production of high-fidelity encoded qubits, called \emph{magic states},
which enable reliable error-corrected computation. We present the first detailed designs of hardware
functional units that implement space-time
optimized \emph{magic-state factories} for surface code error-corrected machines.

Interactions among distant qubits require \emph{surface code braids} (physical pathways on chip) which must be routed. Magic-state factories 
are circuits comprised of a complex set of braids that is more difficult
to route than quantum circuits considered in previous work \cite{javadi2017optimized}. This paper explores the impact of scheduling techniques, such as gate reordering and qubit renaming, and we propose two novel mapping techniques:  braid repulsion and
dipole moment braid rotation.  We combine these techniques with graph partitioning and community detection algorithms, and further introduce
a stitching algorithm for mapping subgraphs onto a physical 
machine.  Our results show a factor of 5.64 reduction in
space-time volume compared to the best-known 
previous designs for magic-state factories.

\end{abstract}

\begin{IEEEkeywords}
Quantum Computing, Quantum Error Correction, Surface Code, Magic State Distillation
\end{IEEEkeywords}

\section{Introduction}\label{sec:Introduction}
Quantum computers of intermediate scale are now becoming a reality. While recent efforts have focused on building Noisy Intermediate-Scale
Quantum (NISQ) computers without error correction, the long-term goal is to build large-scale fault-tolerant machines \cite{preskill2018quantum}. In fault-tolerant machines, typical quantum workloads will be dominated by error correction \cite{tannu2017taming}. On machines implementing \emph{surface code} error correction, fault-tolerant operations known as \emph{magic-state distillation} will make up the majority of the overhead. The problem of achieving effective magic-state distillation is two-fold: 1) useful quantum applications are dominated by magic-state distillation, and 2) their support is extremely expensive in both physical area and latency overhead. The innovations in this paper address the largest obstacle facing large-scale quantum computation. 

Magic-state distillation requires the preparation (i.e. \emph{distillation}) of high-fidelity logical qubits in a particular state, which can enable the execution of fault-tolerant instructions. These states require expensive, iterative refinement in order to maintain the reliability of the entire device.


This work proposes optimizations for the architectural functional units (i.e. ``factories'') to generate magic states. Using a realistic resource overhead model, we introduce optimization techniques that exploit both instruction level scheduling as well as physical qubit mapping algorithms. Our approach analyzes and optimizes, for the first time, fully mapped and scheduled instances of resource state generation units known as multilevel block-code state-distillation circuits. We develop novel technology-independent heuristics based upon physical dipole-moment simulation to guide annealing algorithms aiming to discover optimized qubit register mappings. We use these along with a new combination of conventional compiler methods to exploit structure in the distillation circuitry. Together, these techniques reduce resource overhead (space-time volume) by $5.64$x. We make use of a novel software toolchain that performs end-to-end synthesis of quantum programs from high level expression to an optimized schedule of assembly gate sequences, followed by intelligent physical qubit register allocation, and surface code simulation. 




Our techniques are based on analysis of circuit interaction graphs, where nodes represent qubits and edges represent operations between the endpoints. We show that a combination of graph partitioning-based mapping procedures and dipole-moment driven annealing techniques work well on structured surface code circuits. State distillation circuits can be subdivided cleanly into sets of disjoint planar subgraphs. We find that each of these planar subgraphs can be mapped nearly optimally. The higher level structure of the distillation circuits introduces non-trivial permutation steps between circuit subdivisions as well. We present an algorithm that combines optimized subgraph mappings with a force-directed annealing technique that optimizes the transition between the levels of the circuit. This technique is compared to conventional, global methods that optimize for specific characteristics of the interaction graph. The planar graph extraction and ``stitching" technique outperforms global methods. 

In summary, this paper makes the following contributions:
\begin{itemize}
\item We study the characteristics of two-qubit interactions in surface code error corrected machines, and show strong correlation between circuit latency and the number of edge crossings in the circuit interaction graph.
\item We use this information to develop a heuristic inspired by simulation of molecular dipoles, and show that this can be used to generate low-latency qubit mappings by reducing edge crossings.
\item We exploit the structure of the state distillation circuits to optimize individual rounds of distillation separately, and combine these rounds with optimized permutation networks to generate the lowest resource-overhead implementation of distillation units to date.
\end{itemize}

The rest of the paper is organized as follows: Section \ref{sec:bg} describes quantum computation, surface code error correction, and magic state distillation in more detail. Section \ref{sec:related} describes related work that aims to optimize state distillation. Section \ref{sec:approach} clarifies and summarizes the techniques we use to result in efficient factory circuits. Sections \ref{sec:scheduling} and \ref{sec:mapping} specifically describe the scheduling properties of these circuits and mapping techniques along with heuristics utilized to optimize the procedures. Section \ref{sec:optimize} describes in greater detail the fully optimized algorithm for achieving efficient factory circuits. Section \ref{sec:results} describes the results we obtain. Finally, Sections \ref{sec:future} and \ref{sec:conclusion} discuss future work and conclude.

\section{Background}\label{sec:bg}
\subsection{Basics of Quantum Computation}
\label{subsec:qc}
Quantum computation involves the manipulation of fragile quantum states by operating on quantum bits (qubits). Each qubit is capable of existing in a superposition of two logical states $\ket{0}$ and $\ket{1}$ written as a linear combination $\ket{\psi} = \alpha \ket{0} + \beta \ket{1} $, for complex coefficients $\alpha, \beta$ such that $|\alpha|^2 + |\beta|^2 = 1$. Upon measurement, the qubit state ``collapses" to either $\ket{0}$ or $\ket{1}$. $|\alpha|^2$ and $|\beta|^2$ correspond to the probability of obtaining a $\ket{0}$ or $\ket{1}$ respectively. It is sometimes useful to visualize the state of a single qubit as a vector on the Bloch sphere \cite{bloch1946nuclear,MikenIke}, because we can reinterpret the state $\ket{\psi}$ in its spherical coordinates as $\ket{\psi} = \cos{(\theta/2)}\ket{0} + \exp{(i\phi)}\sin{(\theta/2)}\ket{1}$. Any operations (quantum gates) performed on a single qubit can thus be regarded as rotations by some angle $\varphi$ along some axis $\hat{n}$, denoted as $R_{\hat{n}}(\varphi)$. This work focuses on the phase gate ($S \equiv R_z(\pi/2)$), the T gate ($T \equiv R_z(\pi/4)$), and the most common two-qubit gate called controlled-NOT (CNOT) gate.


Quantum computing systems are commonly characterized by the maximum supportable \textit{space-time volume} of a computation. This is the product of the number of qubits in the system with the number of operations (i.e. timesteps) that can be performed on the system reliably \cite{bishop2017quantum}. Reliable machines can be built in a variety of ways, each of which may result in a different combination of physical qubit count and computation time. To capture this, the space time volume of a computation is a useful metric by which computations and architectural solutions can be compared.

\subsection{Surface Code Error Correction}
\label{subsec:surface}

\begin{figure}[t!]
    \centering
    \includegraphics[width=0.9\linewidth]{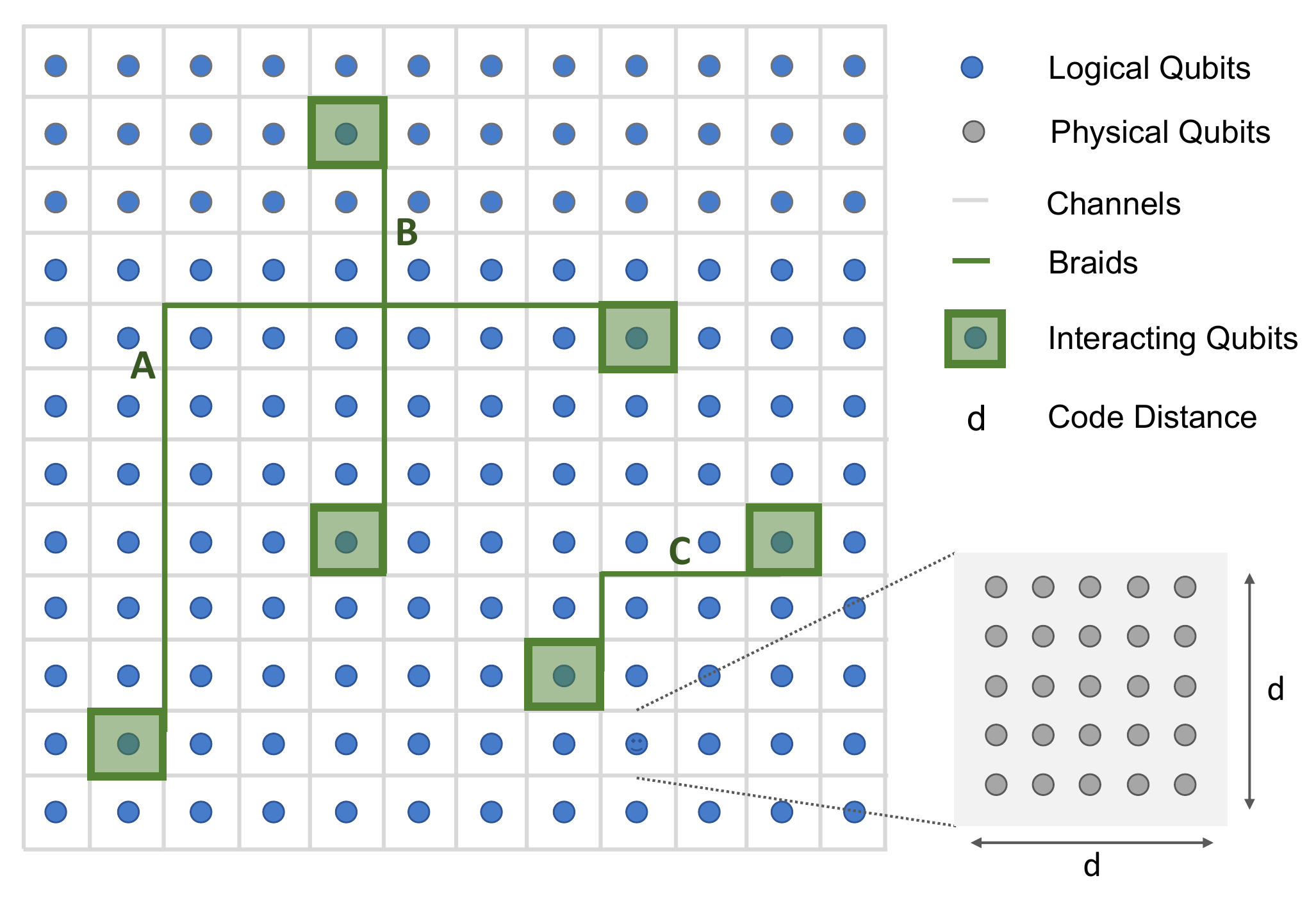}
    \caption{An array of (blue) logical qubits in a quantum processor. Highlighted lines indicate \emph{braids} implementing two qubit interactions. These braids must exist spatially and temporally as pathways between qubits. This introduces communication congestion that depends upon specific architectural designs. Braid $A$ and $B$ are \emph{crossing} braids, which cannot be executed simultaneously, while braid $C$ is isolated and free to execute. Bottom-right inset represents a single logical qubit tile comprised of approximately $d^2$ physical qubit.}
    \label{fig:surfacecode}
\end{figure}

Quantum states decohere over time which can result in performance loss and failure to produce the correct output. In order to maintain the advantage that quantum computation offers while balancing the fragility of quantum states, quantum error correction codes (QECC) are utilized to protect quantum states undergoing a computation. One of the most prominent quantum error correcting codes today is the \emph{surface code} \cite{dennis2002topological,FowlerSurface,jones2012layered}. These codes are a family of quantum error correcting codes that encode logical qubit states into the collective state of a lattice of physical qubits utilizing only nearest neighbor interactions between qubits designated as \textit{data} and \textit{ancilla} qubits. For a comprehensive introduction see an excellent tutorial in \cite{FowlerSurface}. 

An important parameter of the surface code is the \textit{code distance} $d$. The surface code can protect a logical state up to a specific fidelity $P_L$, which scales exponentially with $d$. More precisely, $P_L \sim d(100\epsilon_{in})^{\frac{d+1}{2}}\label{eq:etarg}$, where $\epsilon_{in}$ is the underlying physical error rate of a system \cite{FowlerSurface}. Each logical qubit is made up of approximately $d^2$ physical qubits, as Fig.~\ref{fig:surfacecode} shows.

\subsection{CNOT Braiding}
A \textit{braid}, as illustrated in Fig. \ref{fig:surfacecode}, is a path in the surface code lattice, or an area where the error correction mechanisms have been temporarily disabled and which no other operations are allowed to use. In other words, braids are not allowed to cross. In braiding, a logical qubit is entangled with another if the pathway encloses both qubits, where enclosing means extending a pathway from source qubit to target qubit and then contracting back via a (possibly different) pathway. These paths can extend up to arbitrary length in constant time, by disabling all area covered by the path in the same cycle. 

\subsection{T Gates in Quantum Algorithms}
\label{subsec:algs}
$S$ and $T$ rotation gates are important operations in many useful quantum algorithms, and their error-corrected execution requires magic state resources. When the number of $T$ gates in an application is low, the circuit is in fact able to be efficiently simulated classically \cite{bravyi2016improved}. $T$ gates have been shown to comprise between 25$\%$ and $30\%$ of the instruction stream of useful quantum applications \cite{tannu2017taming}. Others claim even higher percentages for specific application sets, of between $40\%$ and $47\%$ \cite{isailovic2008running}. 



For an estimate of the total number of required $T$ gates in these applications, take as an example the algorithm to estimate the molecular ground state energy of the molecule Fe$_2$S$_2$. It requires approximately $10^4$ iteration steps for ``sufficient" accuracy, each comprised of $7.4 \times 10^6$ rotations \cite{wecker2014gate}. Each of these controlled rotations can be decomposed to sufficient accuracy using approximately 50 $T$ gates per rotation \cite{kliuchnikov2012fast}. All of this combines to yield a total number of $T$ gates of order $10^{12}$. As a result, it is crucial to optimize for the resource overhead required by the execution of $T$ gates at this scale to ensure the successful execution of many important quantum algorithms.

\subsection{T Magic States}
\label{subsec:magic}

$T$ and $S$ gates, while necessary to perform universal quantum computation on the surface code, are costly to implement under surface code. The number of $T$ gates present in an algorithm is the most common metric for assessing how difficult the algorithm is to execute~\cite{Selinger:2013aa,amy2014polynomial}. To achieve fault-tolerance, an ancillary logical qubit must be first prepared in a special state, known as the {\em magic state}~\cite{magic_states}. A distilled magic-state qubit is interacted with the data to achieve the $T$ gate operation, via a probabilistic \emph{injection} circuit involving 2 CNOT braids in expectation. For simplicity, because of their rotation angle relationship, we assume all S gates will be decomposed into two T gates.  

These ancillary quantum states are called magic states because they enable universal quantum computation. Magic states can be prepared using Clifford quantum operations \cite{magic_states}. Since the task of preparing these states is a repetitive process, it has been proposed that an efficient design would dedicate specialized regions of the architecture to their preparation~\cite{steane1997space,jones2012layered}. These {\em magic state factories} are responsible for creating a steady supply of low-error magic states. The error in each produced state is minimized through a process called {\em distillation}~\cite{Bravyi_magic}. 

\subsection{Bravyi-Haah Distillation Protocol}
\label{subsec:BH}
Distillation protocols are circuits that accept as input a number of potentially faulty raw magic states, use some ancillary qubits, and output a smaller number of higher fidelity magic states. The input-output ratio, denoted as $n \rightarrow k$, assesses the efficiency of a protocol. This work focuses on a popular, low-overhead distillation protocol known as the Bravyi-Haah distillation protocol~\cite{Bravyi_magic,fowler2013surface}.

To produce $k$ magic states, Bravyi-Haah state distillation circuits take as input $3k+8$ low-fidelity states, use $k+5$ ancillary qubits, and $k$ additional qubits for higher-fidelity output magic states, thus denoted as the $3k+8\rightarrow k$ protocol. The total number of qubits involved in each of such circuit is then $5k+13$, which defines the area cost of the circuit module.

The intuition behind the protocol is to “make good magic states out of bad ones”. Given a number of low-fidelity states, the protocol uses a syndrome measurement technique to verify quality, and discards states that are bad. Then, the circuit will convert the subset of good states into a single qubit state. The output magic states will have a suppression of error, only if the filtering and conversion follows a particular pattern. This is specified by the parity-check matrix in the protocol. Notably, if the input (injected) states are characterized by error rate $\epsilon_{\text{inject}}$, the output state fidelity is improved with this procedure to $(1+3k)\epsilon_{\text{inject}}^2$. Due to the filtering step, the success probability of the protocol is, to first order, given by $1-(8+3k)\epsilon_{\text{inject}} + \cdots$.

\subsection{Block Codes}
\label{subsec:block}

\begin{figure}[t!]
    \centering
    \includegraphics[width=\linewidth]{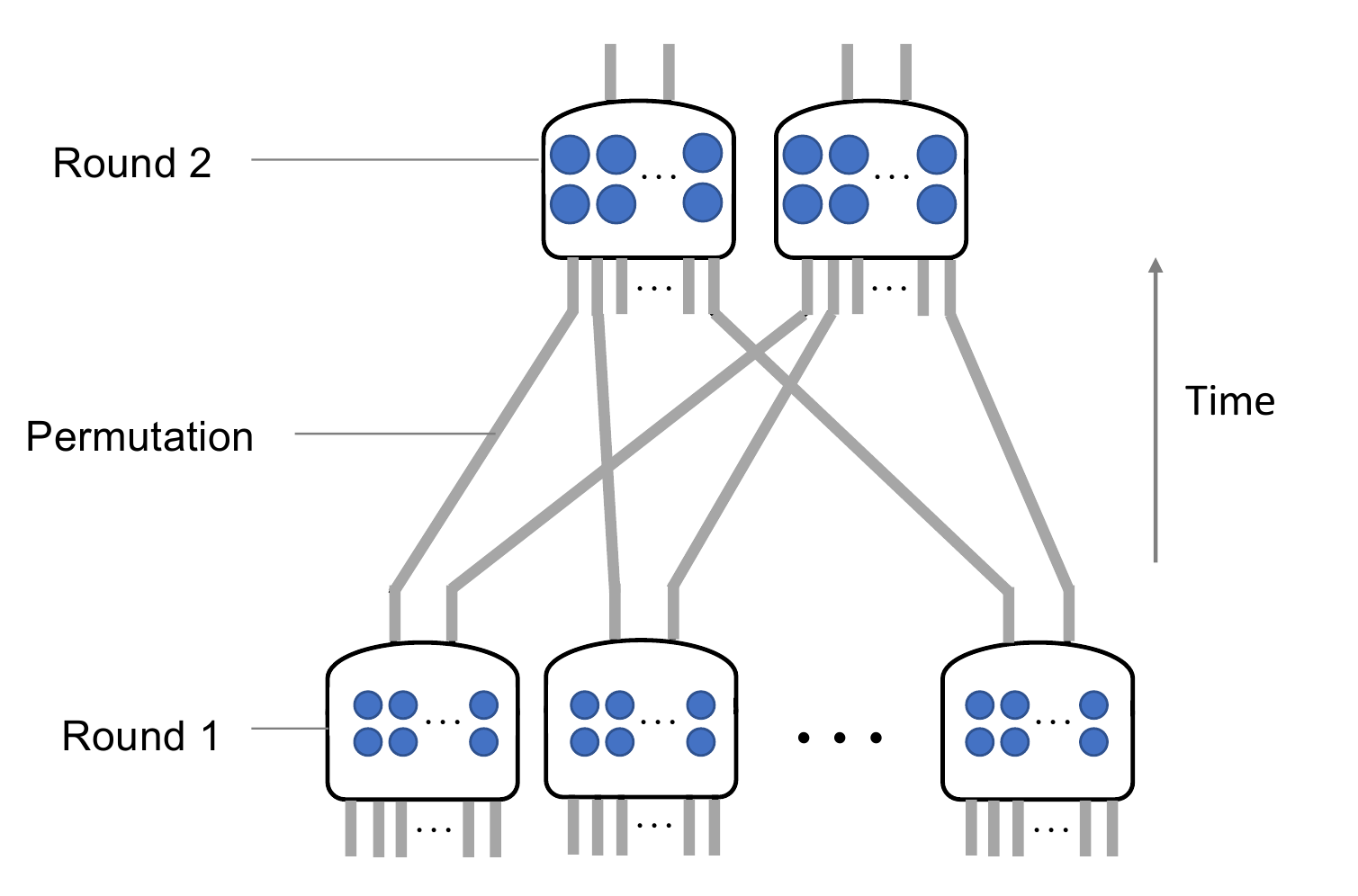}
    \caption{The recursive structure of the block code protocol. Each block represents a circuit for Bravyi-Haah $(3k+8)\rightarrow k$ protocol. Lines indicate the magic state qubits being distilled, and dots indicates the extra $k+5$ ancillary qubits used, totaling to $5k+13$. This figure shows an example of 2-level block code with $k=2$. So this protocol takes as input $(3k+8)^2=14^2$ states, and outputs $k^2=4$ states with higher fidelity. The qubits (dots) in round 2 are drawn at bigger size, indicating the larger code distance $d$ required to encode the logical qubits, since they have lower error rate than in the previous round \cite{campbell}.}
    \label{fig:block_pic}
\end{figure}
Magic state distillation circuits operate iteratively and hierarchically. Often one iteration of the distillation procedure is not enough to achieve the desired logical error rate for a given program. In these cases, squaring the input error rate will not achieve the required logical error rate to execute the program. Instead, we can \textit{recursively} apply the Bravyi-Haah circuit a number $\ell$ times, in order to achieve the desired error rate \cite{jones2013multilevel}. Constructing high fidelity states in this fashion is known as \emph{block code} state distillation.

As Fig. \ref{fig:block_pic} illustrates, $\ell$ level implementations of this procedure can be constructed recursively that support $k^{\ell}$ total output states at fidelity $\sim \epsilon_{\text{inject}}^{2^{\ell}}$, while requiring $(3k+8)^{\ell}$ input states. 

The structure of the multi-level block code distillation protocol requires that each module takes in at most one state from each module from the previous round. This is because the magic states produced by one module may have correlated errors. So in order to avoid having correlated error in the inputs to the next round, each magic state from one module must be fed into a different module.


At the end of each individual module, error checking is performed. If the ancillary states show correct measurement results, the procedure was successful. Additional quality checks were proposed by \cite{campbell}, which inserts a checkpoint at the end of each level of the factory. This checkpoint discards groups of modules when it detects failure within any of the modules in a group. 


Within any particular round $r$ of an $\ell$-level magic state factory, the number of physical qubits required to implement that round defines the \textit{space} occupied by the factory during round $r$. Because the output error rates drop each round, the required code distance increases accordingly. By the ``balanced investment'' technique shown in \cite{campbell}, each logical qubit in round $r$ is constructed using $\sim d_r^2$ physical qubits, where each $d_r$ varies with each round. The idea is to use a smaller code distance to encode a logical qubit in earlier rounds of distillation to minimize area overhead.

In general, any particular round $r$ may require several groups of Bravyi-Haah circuit modules. We denote the number of groups and number of modules per group as $g_{r}$ and $m_{r}$ respectively. The number of physical qubits $q_r$ required to implement that round scales exponentially with $\ell - r$ as: $ q_r = m_{r}^{r-1} g_{r}^{\ell - r}(5k+13)d_r^2$. This exponential scaling plays a key role in our mapping techniques.

\section{Related Work}\label{sec:related}
Other work has focused primarily on optimizing the efficiency of the magic state distillation protocol. The original proposal \cite{bravyi2005universal} considered a procedure by which 15 raw input states would be consumed to produce a single higher fidelity output state. Later works \cite{Bravyi_magic,jones2013multilevel,haah2017magic} each explore different realizations of procedures that distill high fidelity magic states, with each procedure optimizing for asymptotic output rate and increasing this rate from the original proposal. These approaches tend to omit overheads related to actual circuit implementations.

Several prior works \cite{fowler2013surface,campbell} have attempted to reduce the circuit depth of an explicit implementation of the Bravyi-Haah distillation circuit, as well as perform a resource estimate by considering the rates at which these factories fail. Specifically, the work \cite{fowler2013surface} by Fowler et al. is used as a baseline in this paper. 

Additionally, several efforts have been made to build compilers and tools to be more precise about resource estimation quantification in topological quantum error corrected systems \cite{paler2012synthesis,paler2015compiler,paler2016synthesis}. These techniques have resulted in tools that are used to compile and schedule arbitrary quantum circuits to topological assembly, and topological braid compaction techniques are used to reduce circuit depth expansions.

Systems level analysis has been performed by two related projects \cite{isailovic2008running,tannu2017taming}, in which the former optimizes the structure of early distillation protocols, and the latter proposes a micro-architectural accelerator to handle large amounts of error correction instructions that exist in fault tolerant machines. 

Surface code braid scheduling costs were analyzed in \cite{javadi2017optimized} using an end-to-end toolflow. The work focused on the resource impact of the choice of different implementation styles of surface code logical qubits. That work provides a toolchain upon which we have built in order to optimize scheduling and mapping procedures, as well as perform circuit simulations. 

Our work introduces the complexity of braid scheduling into the analysis of the structure of the leading state distillation procedures in an attempt to concretize the procedures into real space and time resource costs. The new annealing heuristics (e.g. dipole-moments) developed specifically for this purpose also generalize well to any circuit executing on a fault tolerant machine that uses braiding to perform two-qubit gates.

\section{Our Approach}\label{sec:approach}


\begin{figure}
    \centering
    \includegraphics[width=0.7\linewidth]{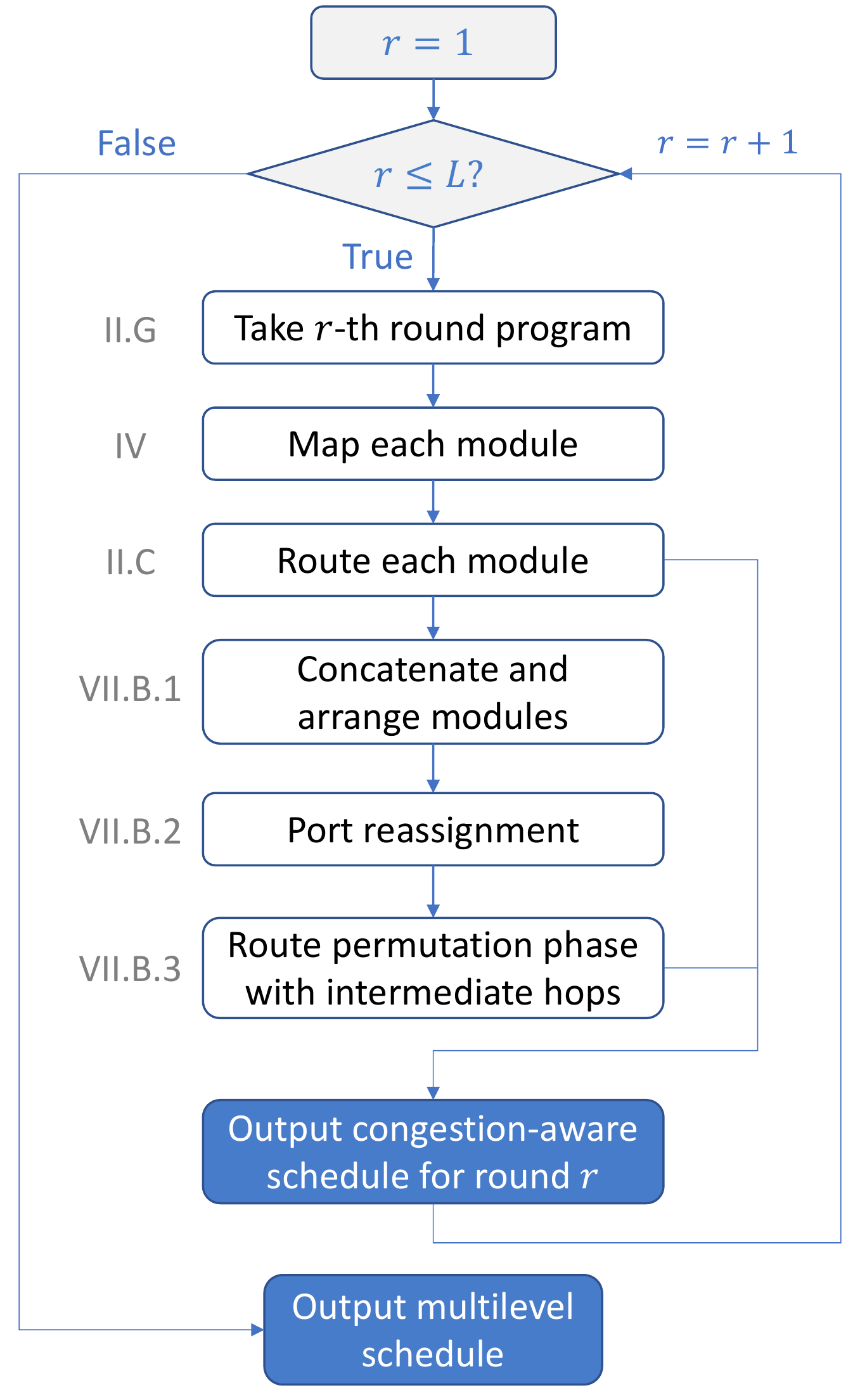}
    \caption{Flow chart for the overall approach, along with the section numbers corresponding to each component.}
    \label{fig:flow}
\end{figure}

\begin{figure*}[t!]
    \centering
    \begin{subfigure}[b]{0.3\linewidth}
        \includegraphics[width=\textwidth,trim=0 -1in 0 0 in]{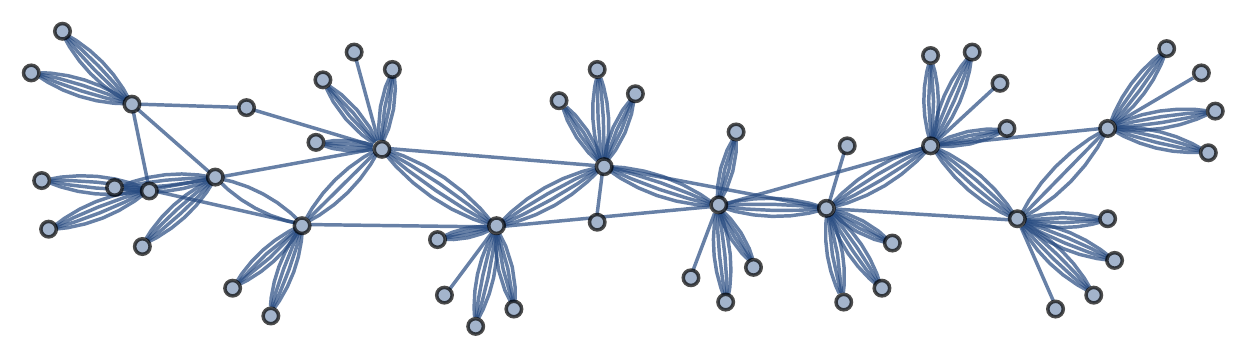}
        \caption{Planar interaction graph of a capacity 8, single level factory}
        \label{fig:k8l1interaction}
    \end{subfigure}
    ~ 
    \begin{subfigure}[b]{0.3\linewidth}
        \includegraphics[width=\textwidth]{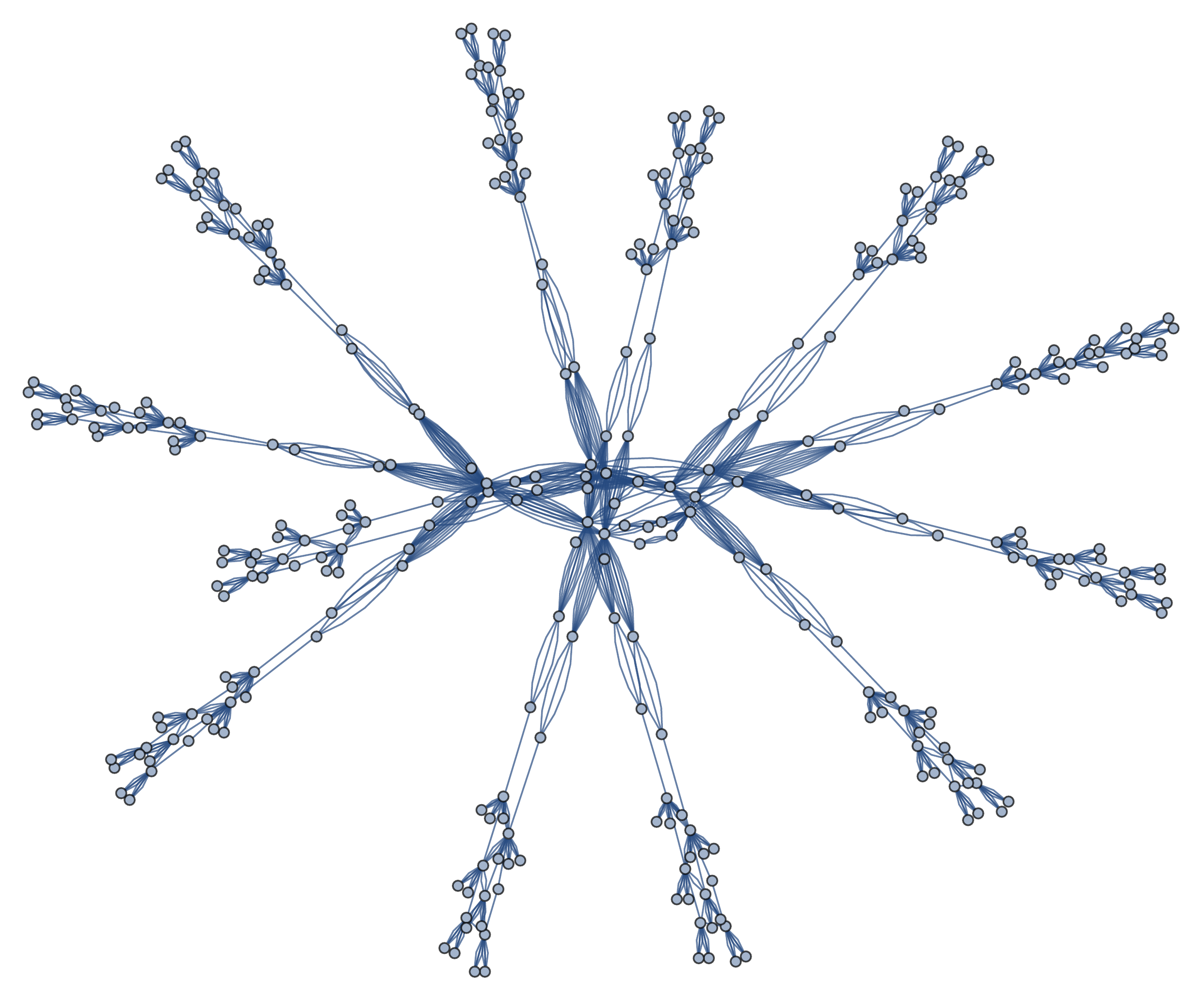}
        \caption{Non-planar interaction graph of a capacity 4, two level factory}
        \label{fig:k4l2interaction}
    \end{subfigure}
    ~ 
    \begin{subfigure}[b]{0.3\linewidth}
        \includegraphics[width=\textwidth]{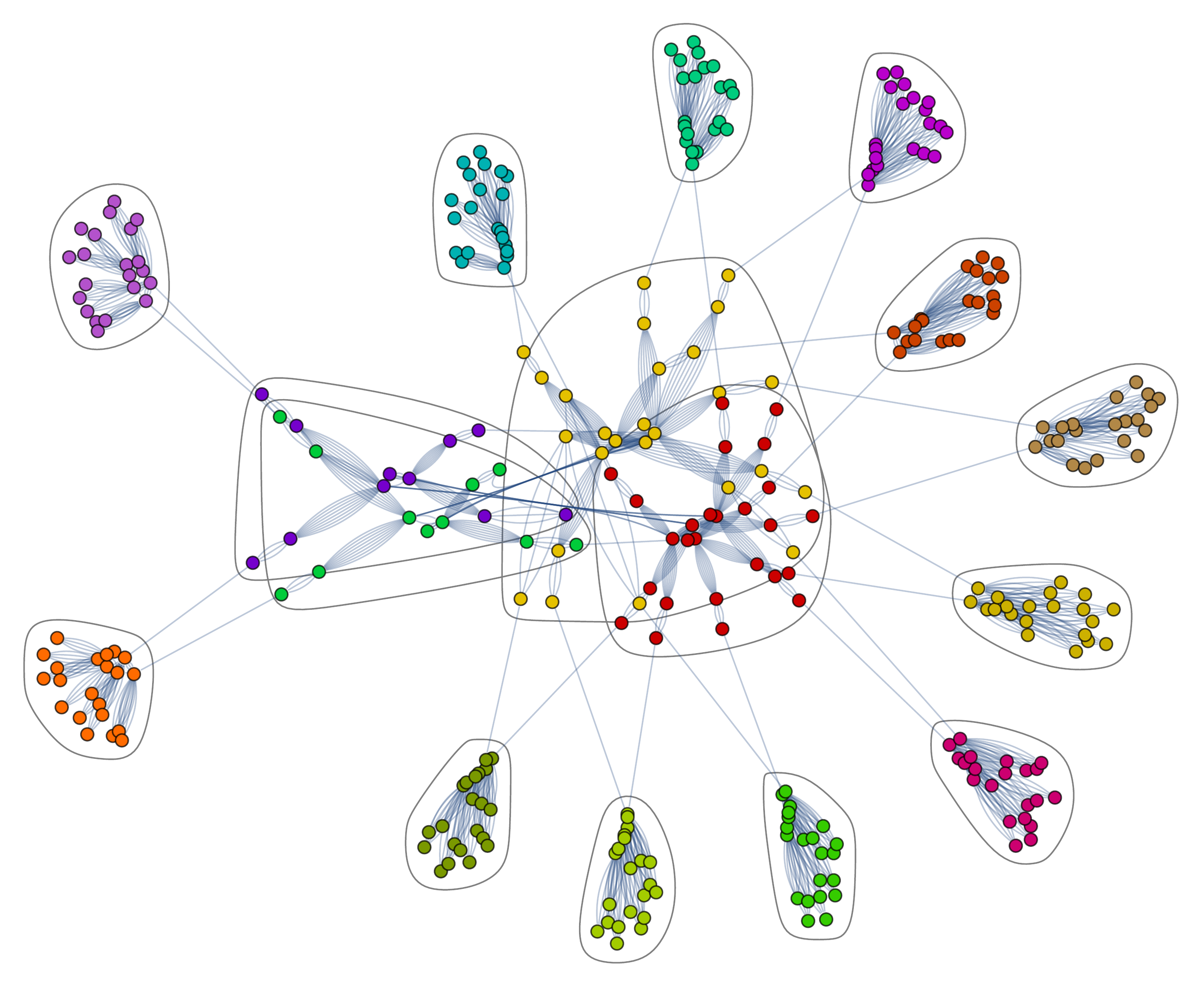}
        \caption{Multi-level factory interaction graph with community structure}
        \label{fig:g42comms}
    \end{subfigure}
    \caption{Interaction graphs of single and two level factories, and community structure of a capacity 4 two level factory. Each vertex represents a distinct logical qubit in the application, and each line represents a required two (or more) qubit operation. (a) shows that the single level distillation circuit has planar interaction graph, so mapping vertices to physical location in quantum processor is relatively simple. Each level in a multi-level factories like (b) have these planar substructures, but the permutation edges between rounds destroy the planarity of the two-level ineraction graph. (c) shows that we can leverage the planarity within each level by exploring community structure of the interaction graph, as shown in Section \ref{sec:mapping}.}
    \label{fig:interactiongraphs}
\end{figure*}

In order to minimize the space-time volume spent on multilevel magic state distillation, our approach takes advantage of the unique characteristics of the state distillation circuitry. We decompose the problem into two aspects -- scheduling gate operations and mapping qubits into 2-D mesh. These two are intertwined, as the schedule determines what pairs of qubits \emph{need} to interact, and mapping influences which subset of them \emph{can} interact in the same cycle. An important tool used to perform these optimizations is the program interaction graph, from which circuit structure can be extracted. In particular, we combine the fact that these state distillation circuits are characterized by natural subdivisions between levels of the factory, with the ability of graph partitioning embedding techniques to nearly-optimally map small planar subgraphs of the program. We exploit this information to design a procedure that decomposes state distillation into components that are independently optimized. The flow chart of the procedure is illustrated in Fig.~\ref{fig:flow}.

Levels of the factory are joined by a specific permutation of the output states exiting from previous rounds of distillation, which appears to impose significant overhead on the whole distillation process. To address this, a force-directed annealing algorithm is used in conjunction with ideas inspired by Valiant intermediate-destination routing for permutation networks \cite{lev1981fast} to reduce the latency of these permutation steps between block code levels. 

The next few sections describe the scheduling and mapping optimizations decoupled from one another, in order to show the specific strengths and weaknesses of each. Section \ref{sec:optimize} then synthesizes these optimizations into a single procedure.
\section{Scheduling}\label{sec:scheduling}

This section describes the impact of instruction level optimizations: gate scheduling and qubit reuse. A \emph{schedule} of a quantum program is a sequence of gate operations on logical qubits. The sequence ordering defines \emph{data dependencies} between gates, where a gate $g_1$ depends on $g_0$ if they share a logical qubit and $g_1$ appears later in the schedule sequence than $g_0$.

\subsection{Gate Scheduling}
The impact of gate scheduling can be quite significant in quantum circuits, and many algorithm implementations rely upon the execution of gates in parallel in order to achieve substantial algorithmic speedup. Gate scheduling in quantum algorithms differs from classical instruction scheduling, as gate commutativity introduces another degree of freedom for schedulers to consider. Compared to the field of classical instruction scheduling, quantum gate scheduling has been relatively understudied, with only few systematic approaches being proposed that incorporate these new constraints \cite{gianScheduling}. 

In exploring these effects applied to Bravyi-Haah state distillation circuits, we find that these optimizations are limited in their effectiveness. While intuitively the modularity of the block code construction would allow for early execution of gates arising in late rounds of the distillation procedure, the \textit{checkpoints} required to implement module checking as described in section \ref{subsec:block} limit the magnitude of gate mobility.

The structure of the block code circuitry only allows for a small constant number of gates to by executed early, outside of the rounds from which they originate. Because of this, the maximum critical path extension by the introduction of a \emph{barrier} preventing gate mobility outside of the originating round is equal to this small constant multiplied by the number of block code iterations. The benefit of inserting a barrier at the end of each round is to create clean divisions between the rounds. As Fig.~\ref{fig:interactiongraphs} shows, the interaction graph for a single round is a planar graph, while this planarity is destroyed as rounds are added. Barriers expose this planarity, making the circuit easier to map. Barriers in these circuits can be inserted by adding a multi-target CNOT operation into the schedule, controlled by an ancilla qubit initialized into a logical $\ket{0}$ state, and targeting all of the qubits that the schedule wishes to constrain.

Additionally, gate scheduling order has significant impacts on network congestion. Scheduling these small constant number of gates early therefore runs the risk of causing congestion with previous round activity. This can in fact extend the circuit latency, even though the circuit has executed gates earlier in the schedule. 

Overall, the insertion of a barrier appears to not significantly alter the schedule of circuit gates. It does, however, change the interaction between the schedule and a particular physical qubit mapping. This relationship will be explored in more detail in Section \ref{sec:optimize}.

\subsection{Qubit Reuse}
We show in this section that an important schedule characteristic of the block protocol to leverage is the hierarchical structure of the distillation circuit. Between two rounds of the procedure, all ancillary qubits will be measured for error checking at the end of the previous round, and reinitialized at the beginning of the next round. This type of data qubit sharing (which we call ``sharing-after-measurement'') is a \emph{false dependency}, because they can be resolved by qubit renaming. Now this naturally leads to the question: (how) should we reuse the qubits between multiple rounds? 

The first approach we explore is to prevent any false sharing of the qubits, at the cost of larger area, by always allocating new data qubits for different rounds. This removes all dependencies due to ancillary qubits, leaving only true dependencies on qubits generated in the previous round. This minimizes execution time at the cost of extra qubits (and space). 

The second approach is to strategically choose which qubits from the previous round to be reused for the next. This approach directly reduces the area needed for the entire factory, at the cost of introducing false dependencies. 

In order to make intelligent decisions on which set of ancillary qubits to reuse, it requires us to have information about the topological mapping of the qubits, since mapping and reuse decisions together significantly influence the congestion overhead of the circuit. We will discuss the subtleties of the tradeoff in more detail later in Section \ref{sec:optimize}.

\section{Mapping}\label{sec:mapping}

\begin{figure}[t!]
\begin{minipage}{\linewidth}
\begin{lstlisting}[language=Scaffold]
// Bravyi-Haah Distillation Circuit with K=8, L=1
#define K 8

module tail(qbit* raw_states, qbit* anc, qbit* out) {
	for (int i = 0; i < K; i++) {
		CNOT ( out[i] , anc[5 + i] );
		injectT ( raw_states[2 * i + 8 + i] , anc[5 + i] );
		CNOT ( anc[5 + i] , anc[4 + i] );
		CNOT ( anc[3 + i] , anc[5 + i] );
		CNOT ( anc[4 + i] , anc[3 + i] );
	}
}

module BravyiHaahModule(qbit* raw_states, qbit* anc, qbit* out) {
 	H ( anc[0] );
 	H ( anc[1] );
 	H ( anc[2] );
 	for (int i = 0; i < K; i++)
 	    H ( out[i] );
 	CNOT ( anc[1] , anc[3] );
 	CNOT ( anc[2] , anc[4] );
 	CXX ( anc[0] , anc , K );
 	tail( raw_states , anc , out );
 	for (int i = 1; i < K + 5; i++)
 		injectT(raw_states[2 * i - 2], anc[i]);
 	CXX ( anc[0] , anc , K + 4 );
 	for (int i = 1; i < K + 5; i++)
 		injectTdag(raw_states[2 * i - 1], anc[i]);
 	MeasX ( anc );
}

/* Single-level circuit requires a single module. 
 * Multi-level circuits would require more modules
 * and barriers in this function. */
module block_code(qbit* raw, qbit* out, qbit* anc) {
 	BravyiHaahModule( raw , anc , out );
}

module main (  ) {
 	qbit raw_states[3 * K + 8];
	qbit out[K];
	qbit anc[K + 5];
 	block_code( raw_states , out , anc );
}
\end{lstlisting}
\end{minipage}
\caption{Example implementation \cite{fowler2013surface, craig_blog} of a single-level Bravyi-Haah distillation circuit generating $K=8$ output magic states, in Scaffold language \cite{Scaffold}. The corresponding interaction graph is illustrated in Fig. \ref{fig:k8l1interaction}. $\mathtt{injectT}$ and $\mathtt{injectTdag}$ implement the probabilistic magic state injection described in \ref{subsec:magic}. $\mathtt{CXX}$ implements a single-control multi-target CNOT gate.}
\label{fig:pseudocode}
\end{figure}

\begin{figure*}[h]
\includegraphics[width=\textwidth, trim=0 0 0 0]{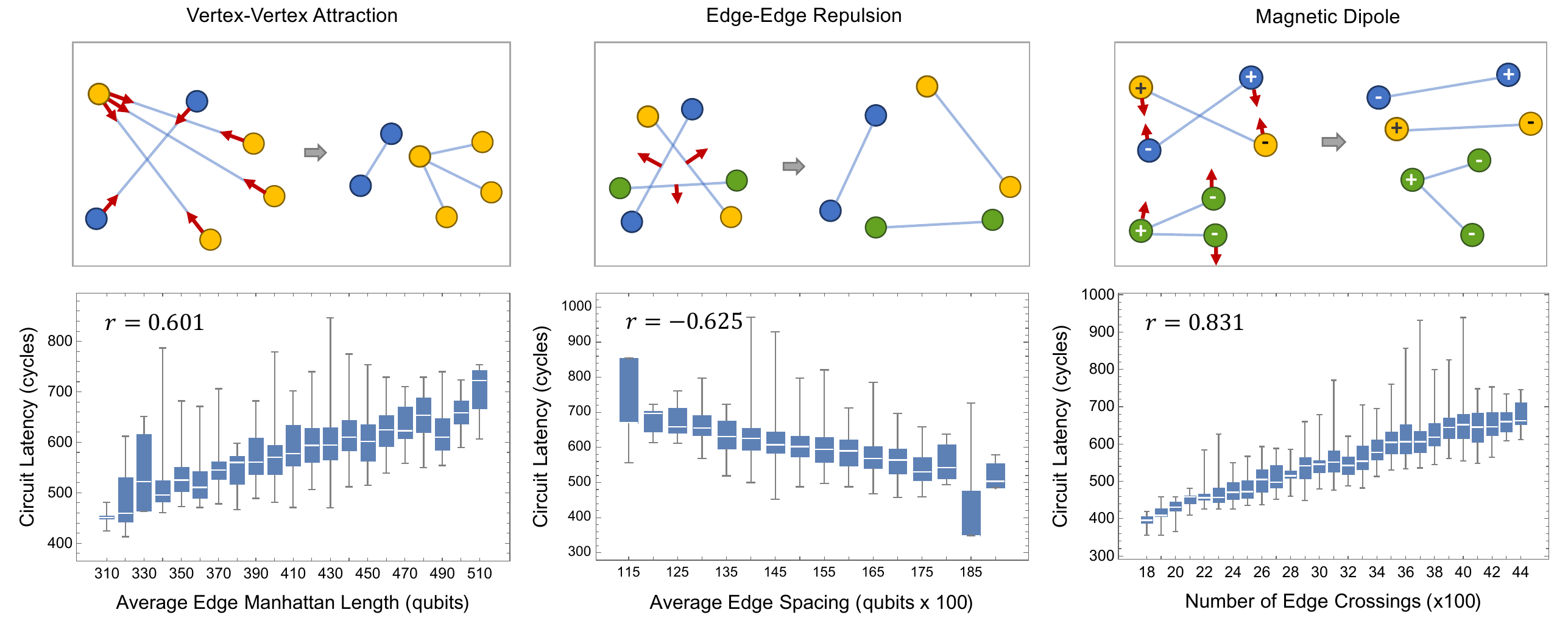}
\caption{Heuristics (top) and metrics (bottom) used in our mapping algorithms, as described in Section \ref{subsec:mapalgo} and \ref{subsec:heuristics} respectively. From left to right, edge length is minimized by vertex-vertex attraction, edge spacing is minimized by repulsion forces on the midpoints of edges, and edge crossings are minimized by applying rotational forces to edges emulating a magnetic dipole moment. For each metric, the correlation coefficient ($r$-value) is calculated across a series of randomized mappings of a distillation circuit, and latency is obtained through simulation, shown in bottom figures. The $r$-values of metrics with latency are $r=0.601$, $-0.625$, and  $0.831$, respectively. The underlying intuition is that shorter edge length, larger edge spacing and fewer edge crossings will result in fewer braid conflicts and shorter overall latency.}
\label{fig:forces}
\end{figure*}

This section describes the impacts of qubit mapping decisions on the overall circuit overhead. Given a schedule we can define a \emph{program interaction graph} as a graph $G = (V,E)$ where $V$ is a set of logical qubits present in the computation, and $E$ is a set of two-qubit interaction gates contained in the program (e.g. CNOT gates). By analyzing this graph, we can perform an optimized \emph{mapping}, which assigns a physical location for each logical qubit $q \in V$. 


 Fig.~\ref{fig:k8l1interaction} and Fig.~\ref{fig:k4l2interaction} depict a single level and a two level factory, respectively, and distinct graph properties are available to analyze for each. The corresponding program that generates Fig.~\ref{fig:k8l1interaction} is shown in Fig.~\ref{fig:pseudocode}. The single level factory is a planar graph. While the two level factory is constructed using many instances of the same single level factory, the requirement for states to be permuted between levels breaks the planarity of the resulting interaction graph. This has significant consequences, and we will leverage them in Section \ref{sec:optimize}.



In order to execute state distillation most efficiently, we must minimize both the area required for the factory as well as the latency required to execute the circuit. Braid operations, while latency insensitive, still cannot overlap with one another. If an overlap is unavoidable, then one operation must stall while the other completes. As a consequence, we aim to minimize the number of these braid ``congestions".

\subsection{Heuristics for Congestion Reduction}\label{subsec:heuristics}

Three common heuristics which we analyze for minimizing network congestion are: edge distance minimization, edge density uniformity, and edge crossing minimization. We see that they each correlate in varying degrees with actual circuit latency overhead for these quantum circuits, as shown in Fig.~\ref{fig:forces}.

\subsubsection{Edge Distance Minimization}
The edge distance of the mapping can be defined as the Euclidean distance between the physical locations of each endpoint of each edge in the interaction graph. Intuitively, in classical systems network latency correlates strongly with these distances, because longer edges require longer duration to execute. As discussed in Section \ref{sec:bg}, for surface code braiding operations, there is no direct correspondence between single edge distance and single edge execution latency. However, longer surface code braids are more likely to overlap than shorter braids simply because they occupy larger area on the network, so minimizing the average braid length may reduce the induced network congestion.

\subsubsection{Edge Density Uniformity} 
When two edges are very close to each other, they are more likely to intersect and cause congestion. Ideally, we would like to maximize the spacing between the edges and distribute them on the network as spread-out and uniformly as possible. This edge-edge repulsion heuristic therefore aims to maximize the spacing between braid operations across the machine.

\subsubsection{Edge Crossings Minimization}
We define an edge crossing in a mapping as two pairs of endpoints that intersect in their geodesic paths, once their endpoint qubits have been mapped. These crossings can indicate network congestion, as the simultaneous execution of two crossing braids could attempt to utilize the same resources on the network. While the edge crossing metric is tightly correlated with routing congestion, minimizing it has been shown to be NP-hard and computationally expensive \cite{garey1983crossing}. An edge crossing in a mapping also does not exactly correspond to induced network congestion, as more sophisticated routing algorithms can in some instances still perform these braids in parallel \cite{chuzhoy2011graph}. Some algorithms exist to produce crossing-free mappings of planar interaction graphs, though these typically pay a high area cost to do so \cite{schnyder1990embedding}.

Fig.~\ref{fig:forces} summarizes the correlation of each of these three metrics to surface code circuit latency.

\subsection{Mapping Algorithms}\label{subsec:mapalgo}
With these metrics in mind, we explore two procedures designed to optimize mappings. First, we employ a local, force-directed annealing optimization technique designed to transform the optimized mappings of Fowler et al. \cite{fowler2013surface} discussed in Section \ref{sec:related}, specifically targeting optimization of the aforementioned metrics. Next, we compare this to a mapping procedure based upon recursive graph partitioning and grid bisection embedding. 

\subsubsection{Force-Directed Annealing}
\label{subsec:forcedirected}
The full force-directed (FD) procedure consists of iteratively calculating cumulative forces and moving vertices according to these forces. Vertex-vertex attraction, edge-edge repulsion, and magnetic dipole edge rotation are used to calculate a set of forces incident upon each vertex of the graph. Once this is complete, the annealing procedure begins to move vertices through the mapping along a pathway directed by the net force calculation. A cost metric determines whether or not to complete a vertex move, as a function of the combination of average edge length, average edge spacing, and number of edge crossings. The algorithm iteratively calculates and transforms an input mapping according to these force calculations, until convergence in a local minima occurs. At this point, the algorithm alternates between higher level \emph{community} structure optimizations that either repulse all nodes within distinct communities away from one another, or attract all nodes within a single community together, which breaks the mapping out of the local minimum that it has converged to. This procedure is repeated until reaching a pre-specified maximum number of iterations.

Within an interaction graph, subsets of qubits may interact more closely than others. These groups of qubits can be detected by performing \emph{community detection analysis} on an interaction graph, including random walks, edge betweenness, spectral analysis of graph matrices, and others \cite{donath1973lower,girvan2002community,fiedler1973algebraic,hughes1995random,blondel2008fast,duch2005community}. By detecting these structures, we can find embeddings that preserve locality for qubits that are members of the same community, thereby reducing the average edge distance of the mapping and localizing the congestion caused by subsets of the qubits. 

\emph{Edge Distance:} To minimize the overall edge distance of the mapping, the procedure calculates the \emph{centroid} of each vertex by calculating the effective ``center of mass" of the neighborhood subgraph induced by this vertex, i.e. the subgraph containing only the vertices that are connected to this vertex, along with the corresponding edges. The center location of this set is calculated by averaging the locations of all of the neighbors, and this is assigned as the centroid for this vertex. This creates an attractive force on this vertex that is proportional in magnitude to the distance between the vertex and the centroid, as shown in the top-left panel in Fig. \ref{fig:forces}. Forces of this type are standard in graph drawing techniques \cite{hu2005efficient}.

%

\emph{Edge Density:} In an attempt to optimize and uniformly distribute the edge density of the mapping, repulsion forces are defined between each pair of distinct edges on the graph. Specifically, for each pair of edges, a repulsion force is created on the endpoints of magnitude inversely proportional to the square of the distance between the midpoints of the edges. This force law is reflected in many typical graph drawing techniques as well, that aim to uniformly distribute graph vertices and edges \cite{lin2012new,fruchterman1991graph}.

\emph{Edge Crossings:} Even though directly minimizing edge crossings in a graph is in general a difficult task to perform, we can approximate it by modeling each edge as a magnetic dipole moment, and the rotational forces applied on each edge will prefer (anti-)parallel orientations over intersecting ones, as shown in Fig. \ref{fig:forces}. North and south poles are assigned to every vertex in the graph, and attractive forces are created between opposing poles, while repulsive forces are added between identical poles. The assignment of the poles is done by producing a 2-coloring of the interaction graph. Notice that the graph is not always 2-colorable, and it usually is not. However, within each time step in the schedule, a vertex (qubit) can have degree at most 2, and is always acyclic. This is because we have a schedule that contains only 2-qubit gates and single-control multi-target CNOTs. Any two gates cannot be performed on the same qubit simultaneously, and the multi-target CNOTs will look like a vertex-disjoint path.

\begin{figure}[t]
    \centering
    \begin{subfigure}[b]{0.48\linewidth}
        \includegraphics[width=\textwidth]{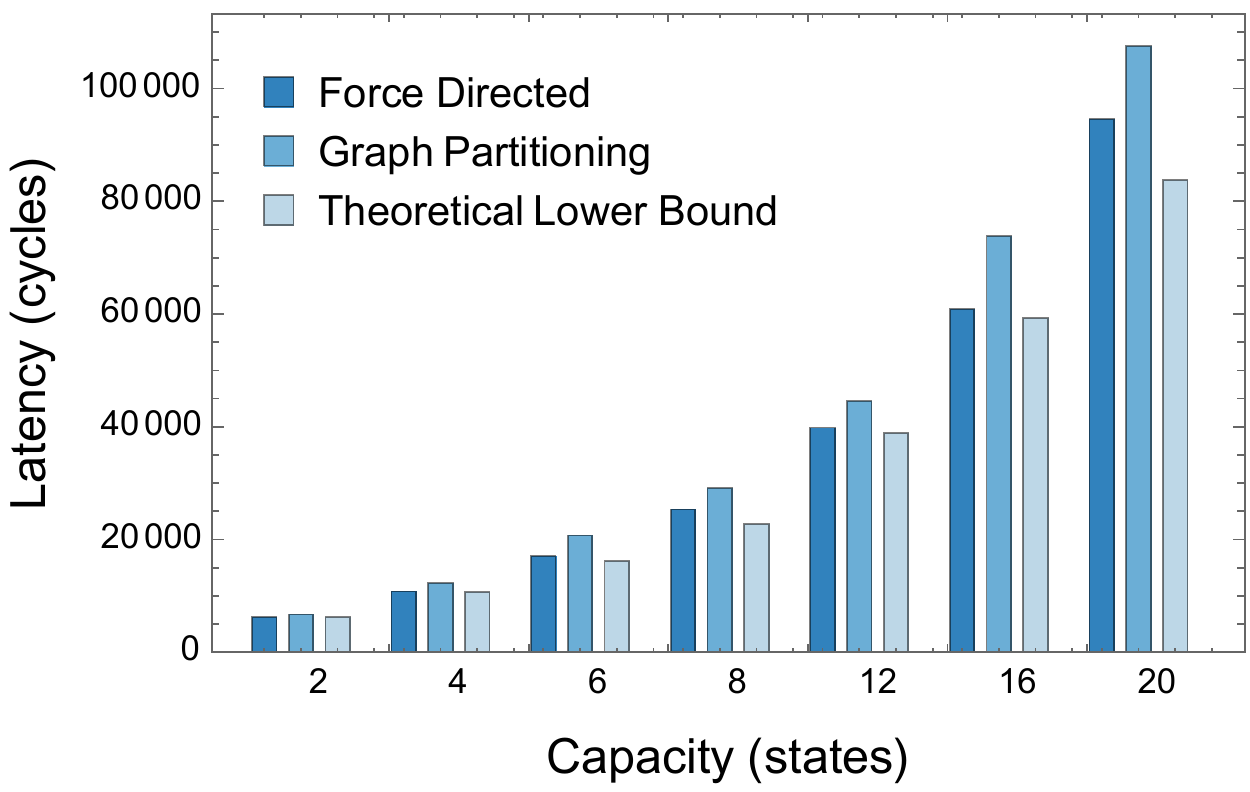}
        \caption{In single-level factories, both techniques can nearly optimally execute these circuits, even as capacity increases.}
        \label{fig:l1gap}
    \end{subfigure}
    ~ 
    \begin{subfigure}[b]{0.48\linewidth}
        \includegraphics[width=\textwidth]{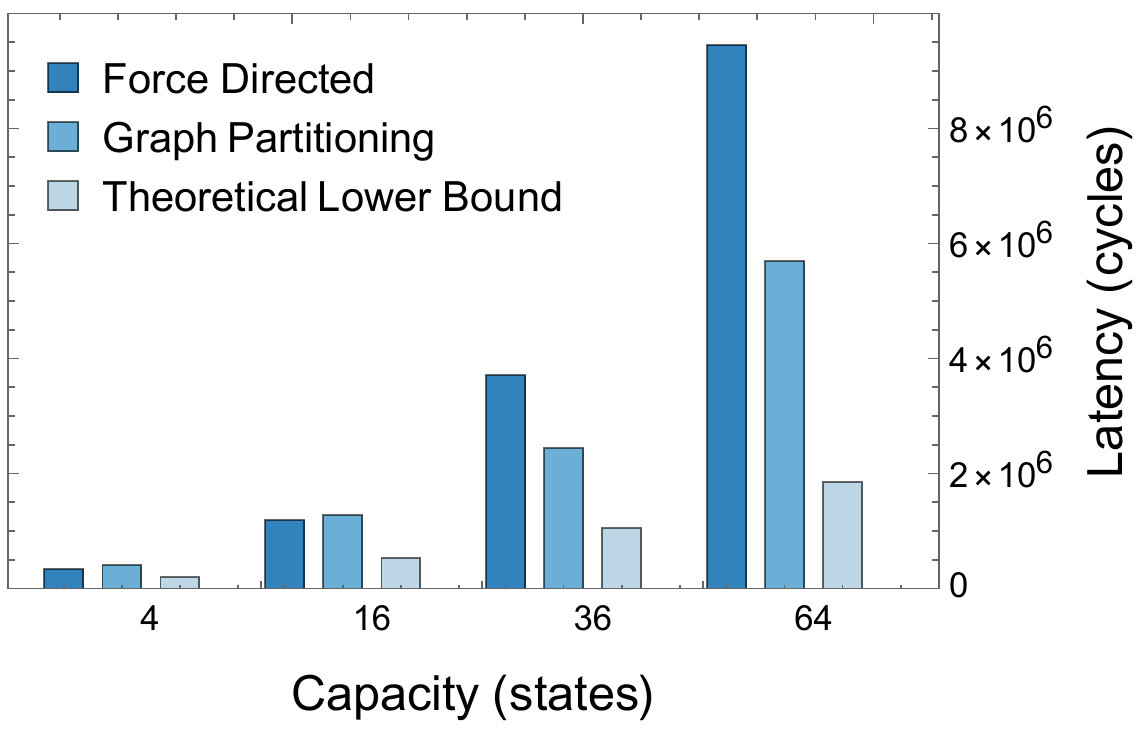}
        \caption{In two-level factories, the difference between the theoretical lower bound and the attained circuit latencies widens.}
        \label{fig:l2gap}
    \end{subfigure}
    \caption{Overall circuit latency obtained by graph partitioning embedding on single and two level distillation factories. Theoretical lower bounds are calculated by the critical path length of the circuits, and may not be physically achievable.}
    \label{fig:fact_time}
\end{figure}

\emph{Community Structure Optimizations:} To respect the proximity of the vertices in a detected community, we break up our procedure into two parts: firstly, impose a repulsion force between two communities such that they do not intersect and are well separated spatially; secondly, if one community has been broken up into individual components/clusters, we join the clusters by exerting attracting forces on the clusters. In particular, we use the KMeans clustering algorithm \cite{kanungo2002efficient, arthur2007k} to pinpoint the centroid of each cluster within a community and use them determine the scale of attraction force for joining them.

\subsubsection{Recursive Graph Partitioning}
\label{subsec:graphpartitioning}
To compare against the local force-directed annealing approach, we also analyzed the performance of a global grid embedding technique based upon graph partitioning (GP)  \cite{kernighan1970efficient,barnes1982algorithm,karypis2000multilevel}. In particular, we utilized a recursive bisectioning technique that contracts vertices according to a heavy edge matching on the interaction graph, and makes a minimum cut on the contracted graph. This is followed by an expanding procedure in which the cut is adjusted to account for small discrepancies in the original coarsening \cite{karypis1995metis,pellegrini1996scotch}. 
Each bisection made in the interaction graph is matched by a bisection made on the grid into which logical qubits are being mapped. The recursive procedure ultimately assigns nodes to partitions in the grid that correspond to partitions in the original interaction graph.


The primary difference between these two techniques is that the former force-directed approach makes a series of local transformations to a mapping to optimize the metrics, while the graph partitioning approach can globally optimize the metrics directly.


\subsubsection{Scalability Analysis}
We can now compare the computational complexity of the two graph optimization procedures. Suppose we have an interaction graph of $n$ vertices and $m$ edges. Each iteration of the force-directed annealing procedure consists of three steps, vertex attraction, edge repulsion, and dipole moment rotation. In the worst case, the attraction forces are computed along each edge in $\mathcal{O}(m)$ time; the repulsion force computation requires $\mathcal{O}(m^2)$ time; rotations are calculated first by a DFS-style graph coloring and then by forces between vertices with $\mathcal{O}(n^2)$. 

Graph partitioning requires recursively finding minimum weight cut, and partition the graph along the cut. Specifically, it requires $\log_2(n)$ recursive iterations, each of which is a min-cut algorithm on partitions of the graph that requires $\mathcal{O}(n+m)$ time, combining to $\mathcal{O}((n+m)\log_2(n))$ \cite{karypis1995metis}.

\subsubsection{Performance Comparison}
Fig. \ref{fig:l1gap} and \ref{fig:l2gap} indicate that, while both techniques perform well for single level factories, the global technique is much better at optimizing higher level factories. This is likely due to the local nature of the force-directed procedure, which is being used to transform the linear hand-optimized initial mapping of the factory. For higher level factories, this hand-optimized mapping incurs high overheads, and the local optimizations are only able to recover a portion of the performance proportional to the original mapping.

While the global graph partitioning technique works well in comparison with the local procedure, there is a widening performance gap between the resulting mapping and the critical resource volume, as factories grow in capacity and levels. This likely indicates that while the procedure is able to very effectively optimize small planar graphs, it has a more difficult time as the size and complexity of the graphs increase. In fact, single level factories have planar interaction graphs, and graph partitioning is able to optimize the mapping of these graphs nearly up to critical resource volume.

\section{Hierarchical Stitching Method}\label{sec:optimize}

\begin{figure}
    \centering
    \includegraphics[width=0.52\textwidth]{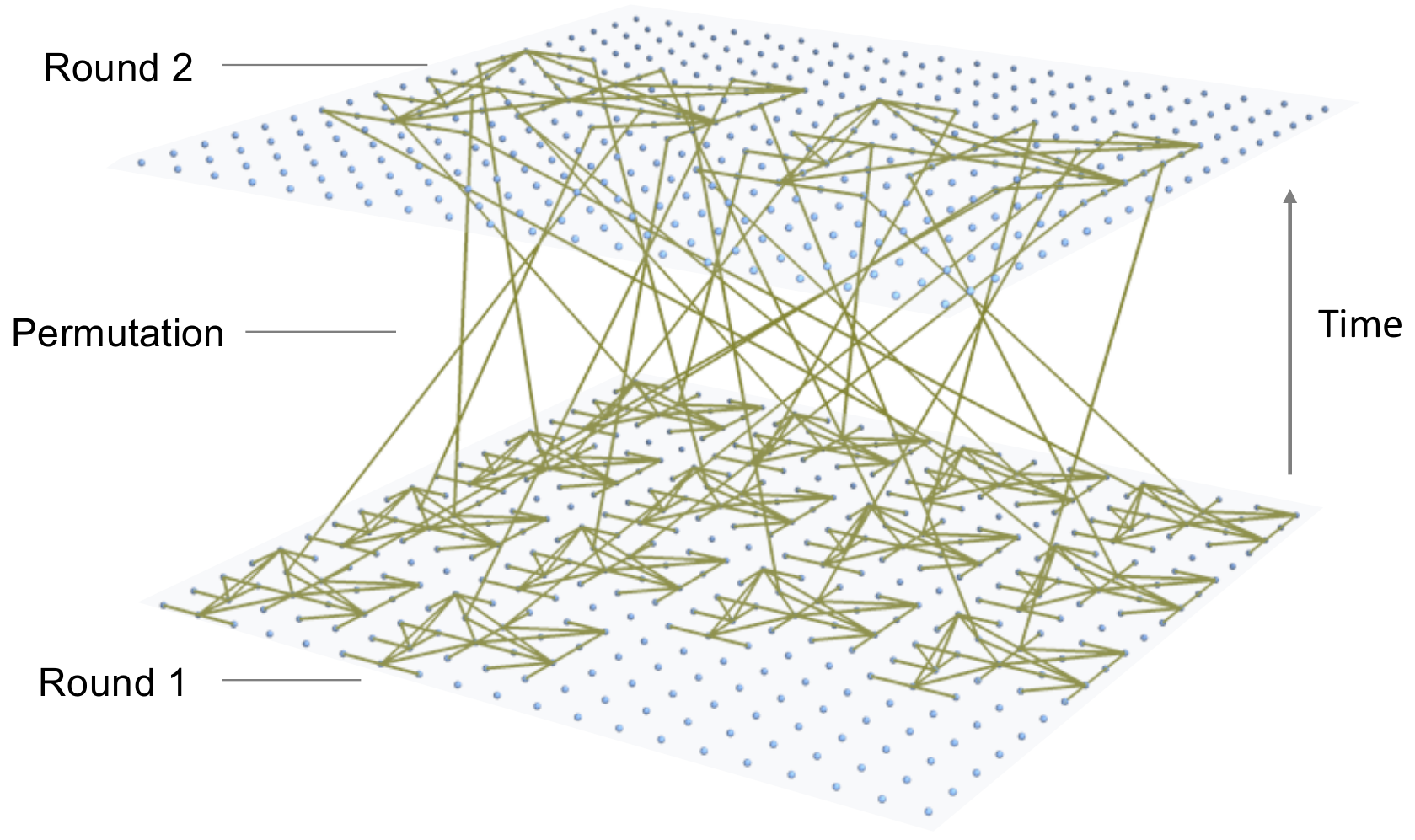}
    \caption{Embedding for a capacity $K=4$, level $L=2$ factory. The stitching procedure optimizes for each round to execute at nearly critical path length in latency, and optimizes for inter-round permutation step with force-directed optimizations.}
    \label{fig:3d}
\end{figure}

We here present the outline of the iterative, synthesized optimization procedure that make use of the scheduling and mapping techniques we established earlier. To take advantage of the facts that most global optimization techniques (such as graph partitioning and force-directed annealing) work well on small planar graphs and that the circuit modules within each round form disjoint planar subgraphs, we develop a stitching scheme, as depicted in Fig.~\ref{fig:3d}, that respects the hierarchical structure and internal symmetry of the multilevel block protocol while simultaneously optimizing for the previously discussed congestion heuristics.


As shown in Fig.~\ref{fig:flow}, we perform optimizations iteratively on the interaction graph. In each iteration, our procedure is decomposed into two phases: (1) \emph{inter-round} optimization that embeds and concatenates each module in the current round, and (2) \emph{intra-round} optimization that stitches permutation edges and arranges modules in the next round.

\subsection{Intra-Round Graph Concatenation}
\label{subsec:intra-stitch}

Starting with the first round of a multilevel factory, we use single-level optimization techniques (such as force-directed annealing or graph partitioning) to nearly optimally embed the \emph{individual} planar modules. They are then \emph{concatenated} together to form a full mapping of the first round of the factory circuitry. The concatenation scheme works well due to the fact that modules in a round do not interact with each other under block code protocol. Notice that putting barrier between rounds enables us to isolate and individually optimize for each round, as discussed in Section \ref{sec:scheduling}. Because the modules in each round of the factory are identical in schedule to those in all other rounds, the optimized graph partitioning embedding does not need to change for each round. 

\begin{figure*}[t!]
    \centering
        \begin{subfigure}{0.24\textwidth}
        \includegraphics[width=\linewidth]{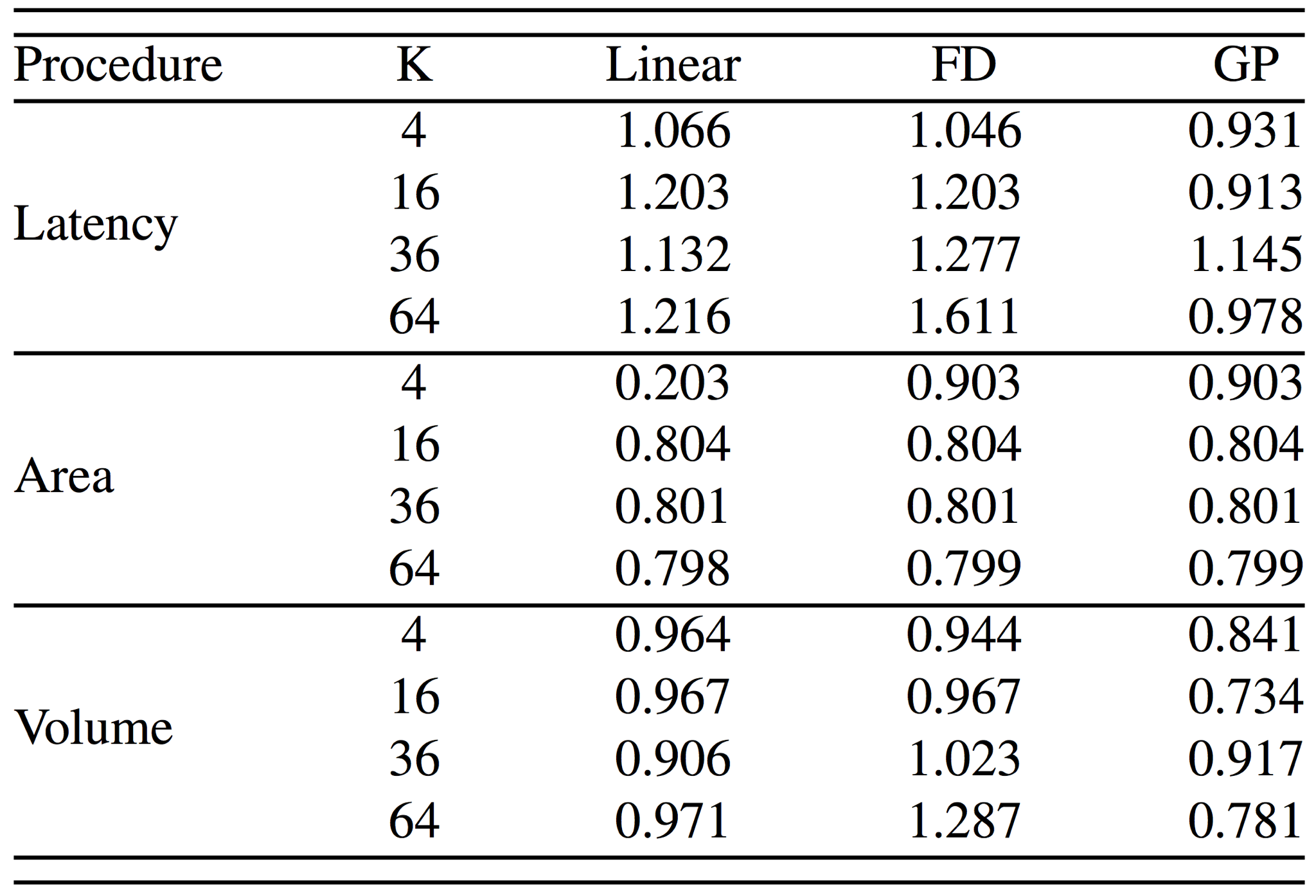}
        \caption{Resource ratios comparing qubit reuse to non-reuse protocols}
        \label{fig:g42comms}
    \end{subfigure}
    ~
    \begin{subfigure}{0.27\textwidth}
        \includegraphics[width=\linewidth, trim=0 -0.5in 0.2in 0]{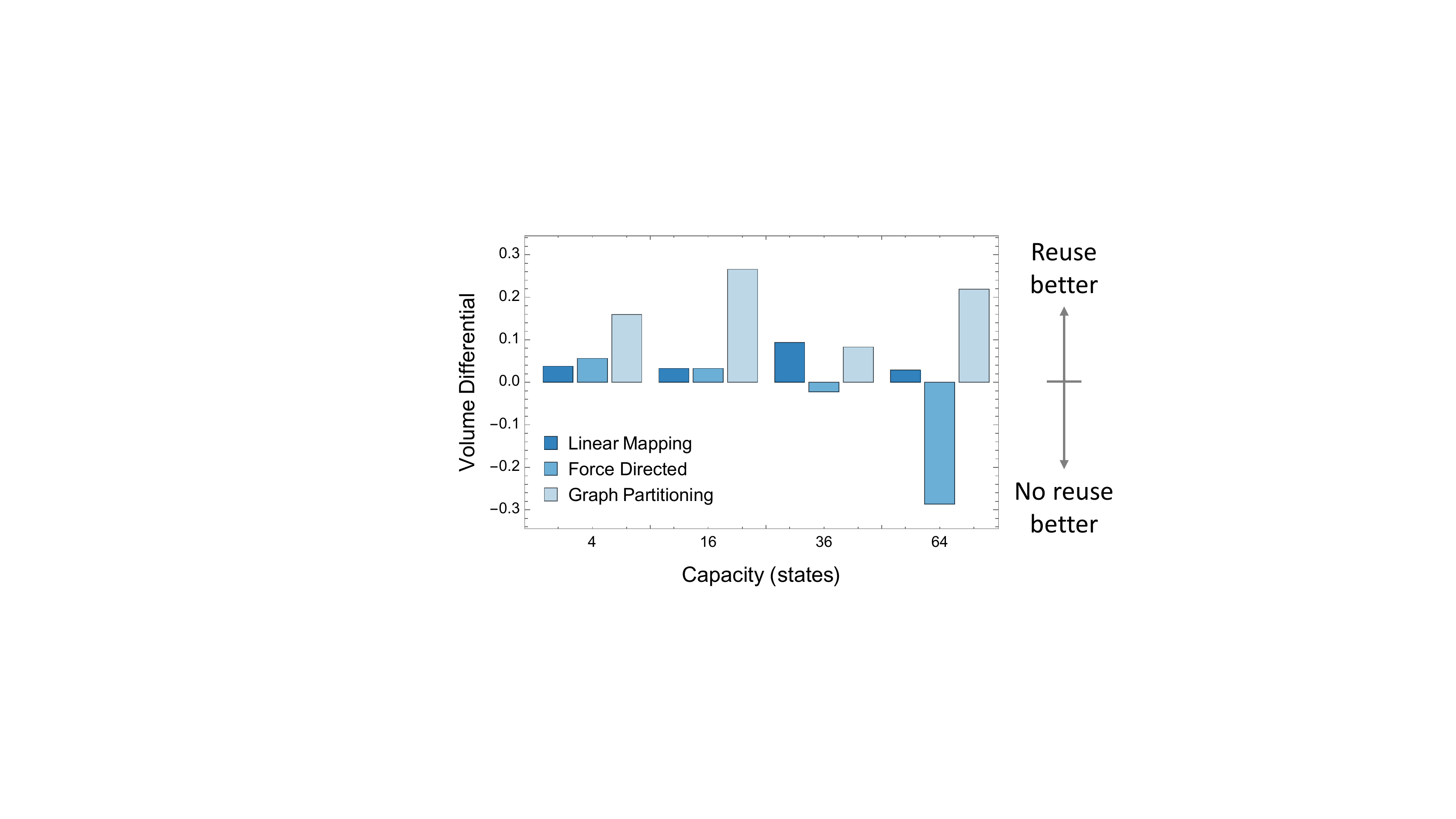}
        \caption{Volume differentials between\\ qubit reuse and non-reuse protocols.}
        \label{fig:k42interaction}
    \end{subfigure}
    ~
    \begin{subfigure}{0.20\textwidth}
        \includegraphics[width=0.8\linewidth, trim=-0.5in 0 1in 0in]{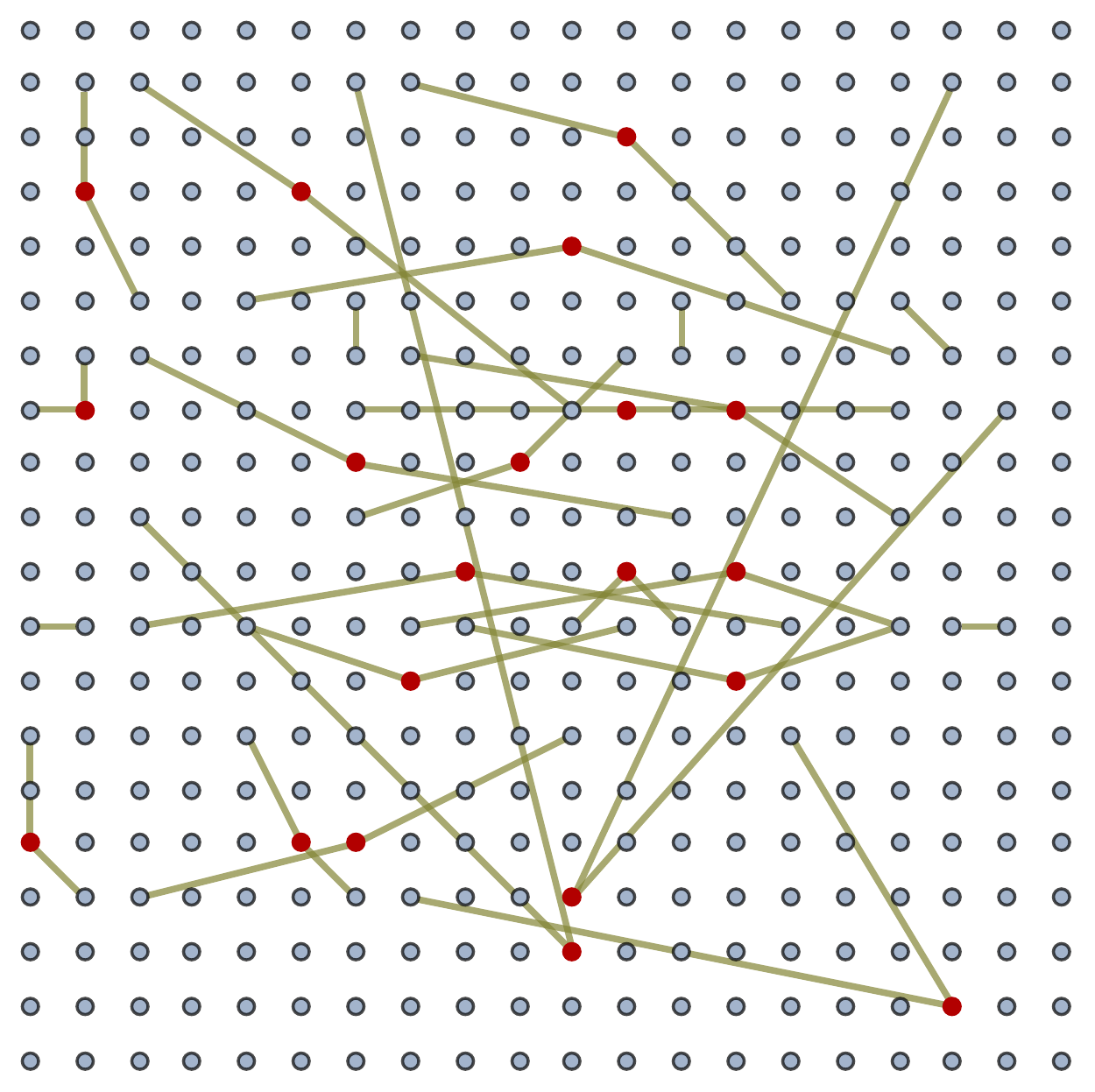}
        \caption{Red dots show optimized intermediate destinations.}
        \label{fig:valiantinteraction}
    \end{subfigure}
    ~
    \begin{subfigure}{0.21\textwidth}
        \includegraphics[width=\linewidth, trim=0 0 0 0in]{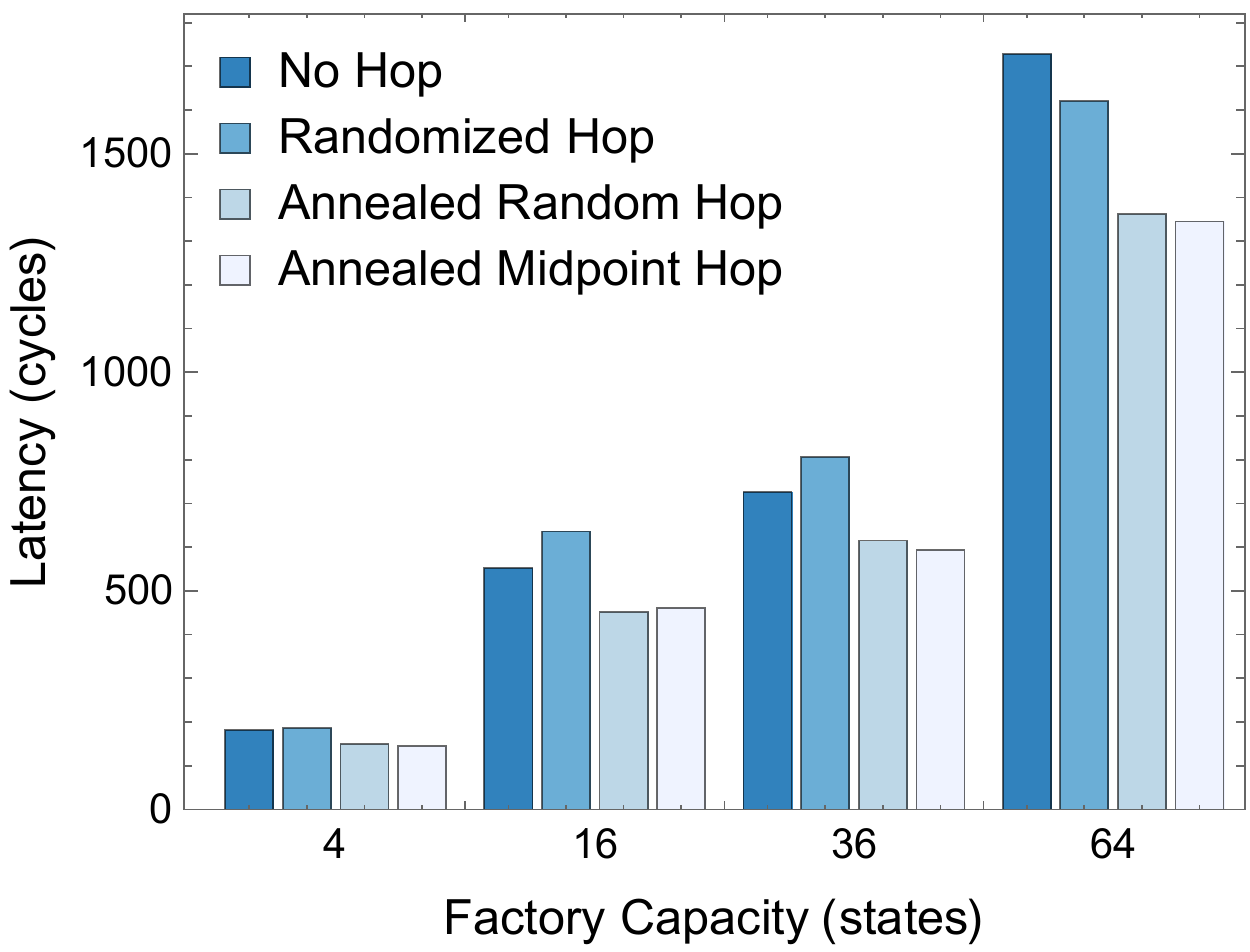}
        \caption{Permutation latencies by no midpoint, Valiant, or annealed midpoints.}
        \label{fig:valiantPlot}
    \end{subfigure}
\caption{(a)-(b): Sensitivity of achievable quantum volumes by different optimization procedures. Shown is the percentage difference of the protocol with reusing (R) or without reusing (NR) qubits: $(\text{NR}-\text{R})/\text{NR}$. Notably, reuse policy is a better for both the linear mapping and graph partitioning techniques, while no-reuse offers more flexibility for force-directed procedure to optimize. (c)-(d): Circuit latency specifically for the inter-round permutation step. Latency is reduced by $1.3$x with Valiant-style intermediate destinations for each interaction, and using force-directed annealing to optimize their locations.}
    \label{fig:valiant}
\end{figure*}

\subsection{Inter-Round Permutation Optimization}
\label{subsec:inter-stitch}

The recursive block code structure requires that the output from lower levels of the factory be permuted and sent to new locations for the subsequent rounds. This can create highly-congested ``permutation steps" in the circuit, where even though each round is scheduled and mapped nearly optimally, the cost to permute the outputs of one round to the inputs of the next round are quite high, as illustrated in the comparison of Fig.~\ref{fig:valiantPlot} with Fig.~\ref{fig:l2latencies}. We therefore present the following sequence of procedures that target the inter-round communication/permutation overhead.

\subsubsection{Qubit Reuse and Module Arrangement}
The permutation edges in between two rounds are due to communications between the \emph{output} qubits from the previous round and the \emph{input} qubits in the next round. Given an optimal layout of the modules/blocks from the previous round, we know where the output states are located. Since all qubits except for the outputs are measured, error-checked, and then reinitialized by the time the next round starts, we can choose which regions of qubits to be reused for the next round, as long as for each module the following constraints are satisfied: (1) do not overlay a module on top of output qubits that are not supposed to be permuted to this particular module (see details about port assignment in \ref{subsubsec:reassign}), and (2) allocate enough qubits required by the code distance as discussed in \ref{subsec:block}. Fig. \ref{fig:g42comms} and \ref{fig:k42interaction} show that reusing qubits benefits the linear and graph partitioned mapping techniques, while force-directed annealing prefers the flexibility added by not reusing qubits.

\subsubsection{Port Reassignment}\label{subsubsec:reassign}
To avoid having correlated error in the inputs to the next round, each module in the next round must gather input states from different modules in the previous round, as shown in \ref{subsec:block}. Suppose one module from the next round wants a magic state from a previous-round module, when there are multiple outputs produced in that module, it does not matter which one you choose. Therefore, it leaves the optimization procedure to decide which output port to use, so as to minimize congestions in the permutation step.

\subsubsection{Intermediate Hop Routing}
Lastly, we employ a variation of the force-directed annealing algorithm from Section \ref{sec:mapping}. Specifically, we introduce \emph{intermediate destinations} between each output state from a prior round and the input state to the next round, as depicted in Fig.~\ref{fig:valiant}. While Valiant routing with randomized intermediate destinations does not increase performance very significantly, we are able to use force-directed annealing based upon edge distance centroids, edge repulsion, and edge rotations in order to move the intermediate destinations into preferable locations.


This synthesized procedure is able to leverage the scheduling techniques of barrier insertion, combined with nearly optimal planar graph embedding performed by recursive graph partitioning, and force-directed annealing to obtain a significant resource reduction over any other optimization procedures.

\section{Results}\label{sec:results}

\begin{figure*}[t!]
    \centering
    \begin{subfigure}[b]{0.23\linewidth}
    \includegraphics[width=\textwidth]{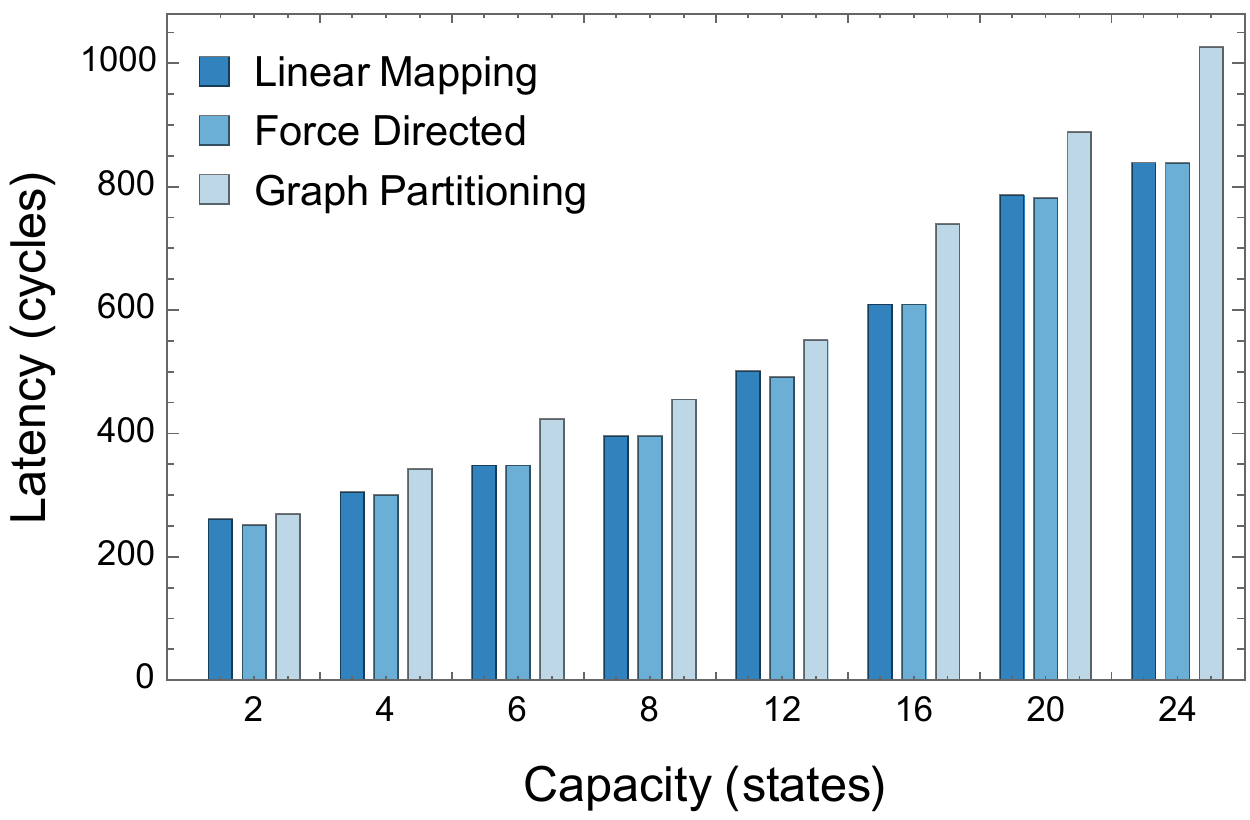}
    \caption{}
    \label{fig:l1latencies}
    \end{subfigure}
    ~
    \begin{subfigure}[b]{0.23\linewidth}
    \includegraphics[width=\textwidth]{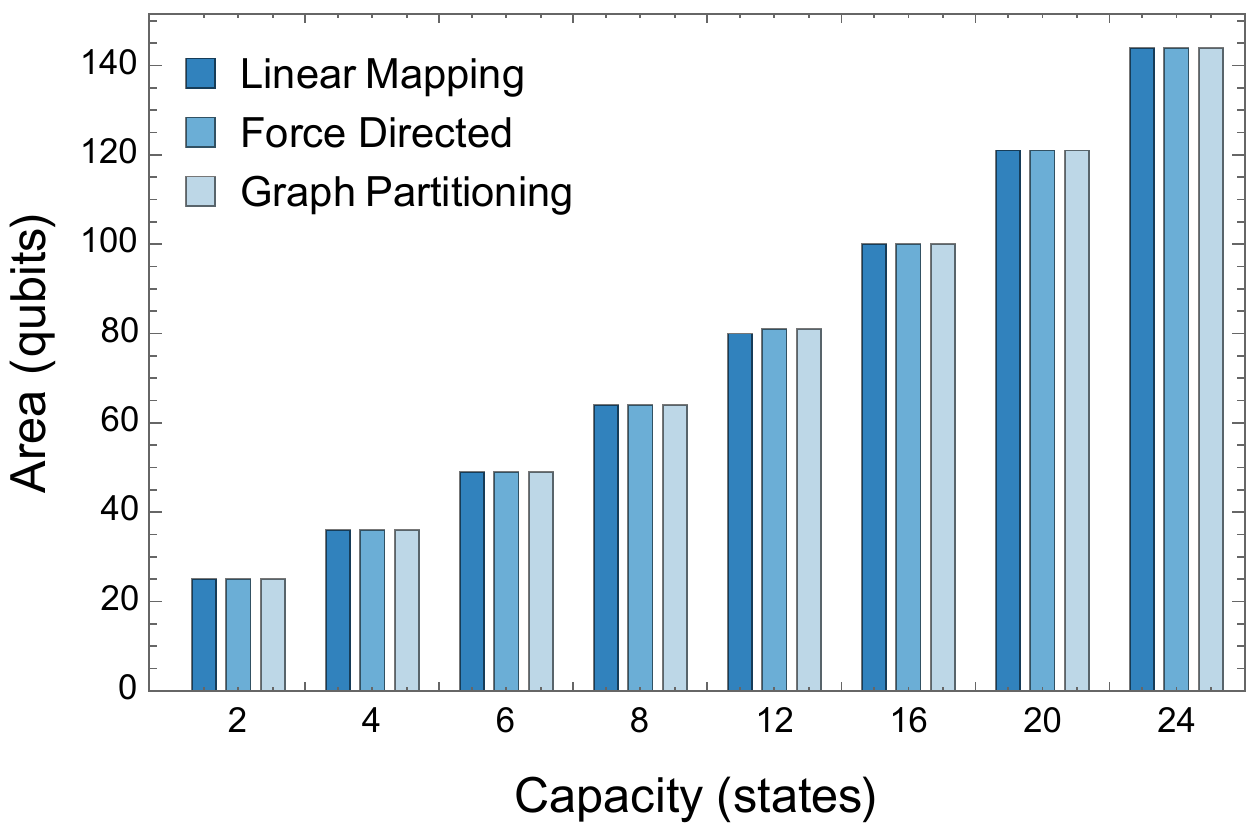}
    \caption{}
    \label{fig:l1areas}
    \end{subfigure}
    ~
    \begin{subfigure}[b]{0.23\linewidth}
    \includegraphics[width=\textwidth]{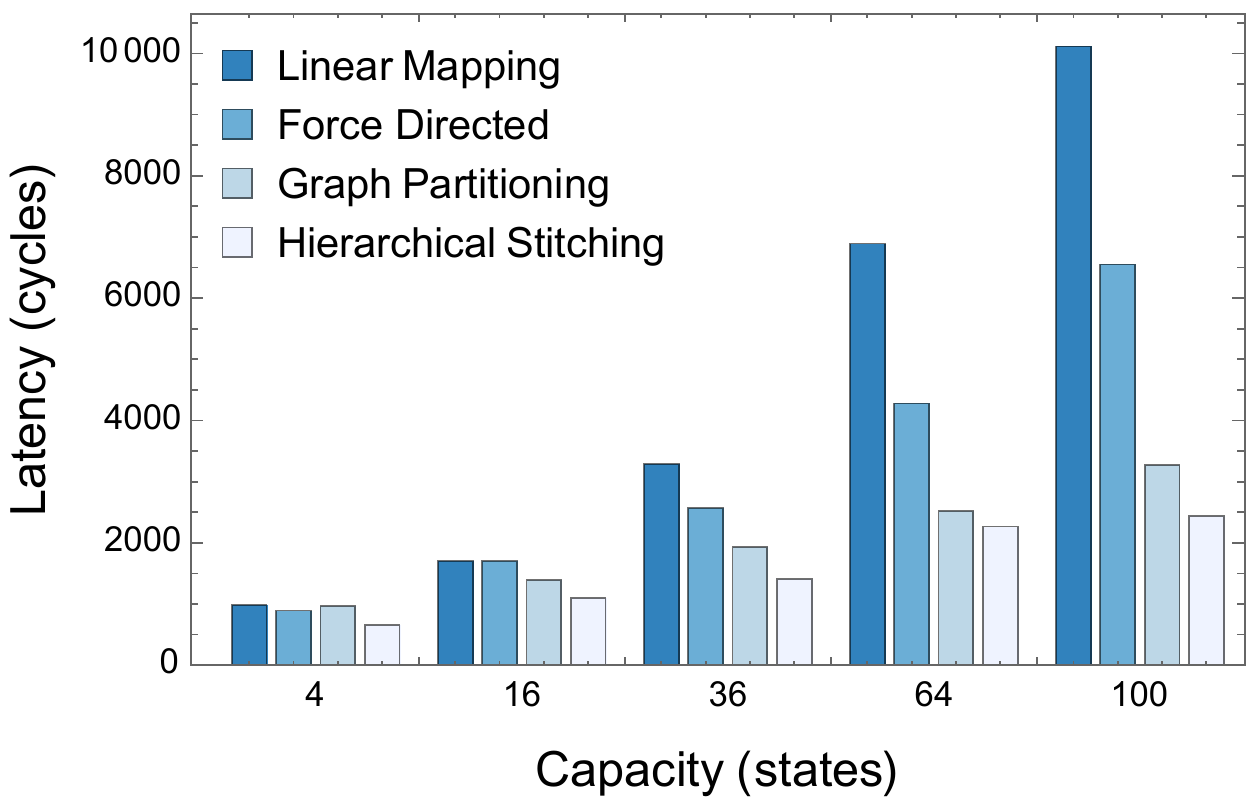}
    \caption{}
    \label{fig:l2latencies}
    \end{subfigure}
    ~
    \begin{subfigure}[b]{0.23\linewidth}
    \includegraphics[width=\textwidth]{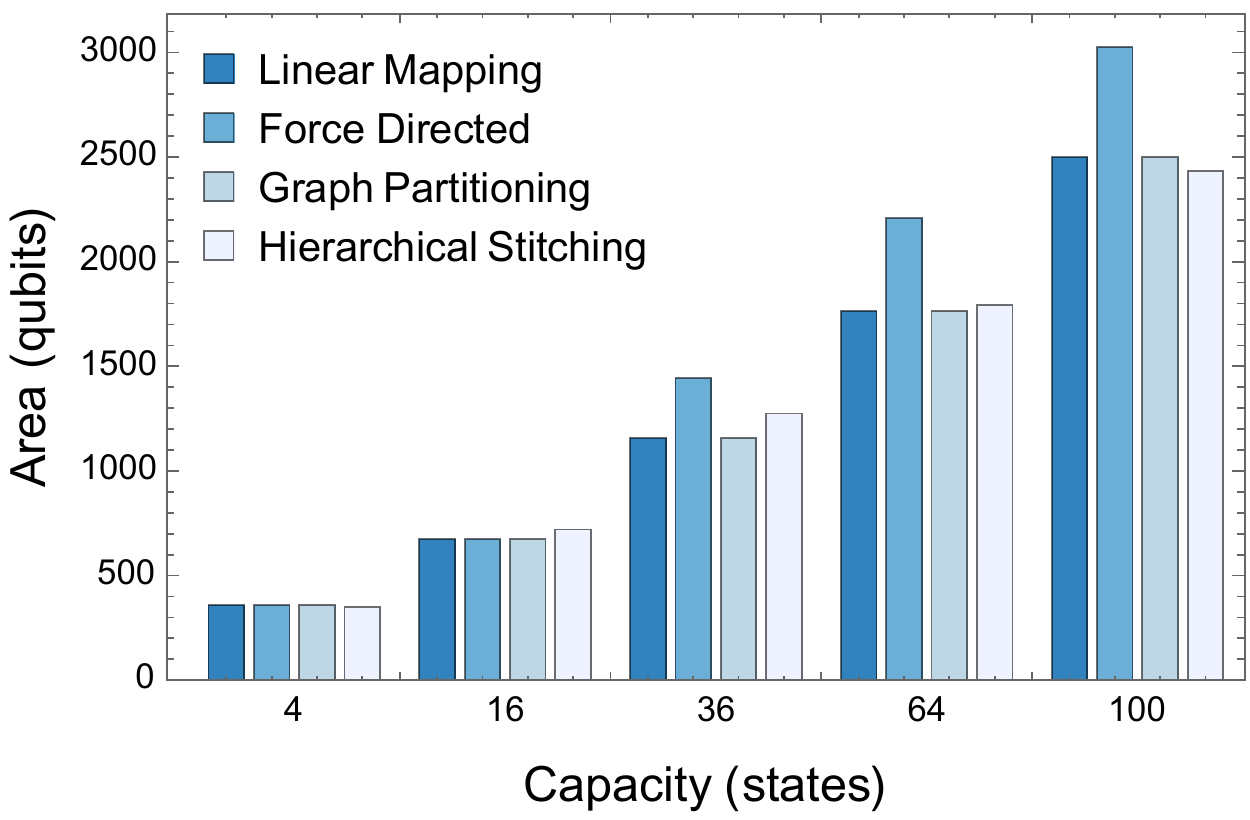}
    \caption{}
    \label{fig:l2areas}
    \end{subfigure}
    ~\\
    \begin{subfigure}[b]{0.33\linewidth}
    \includegraphics[width=\textwidth]{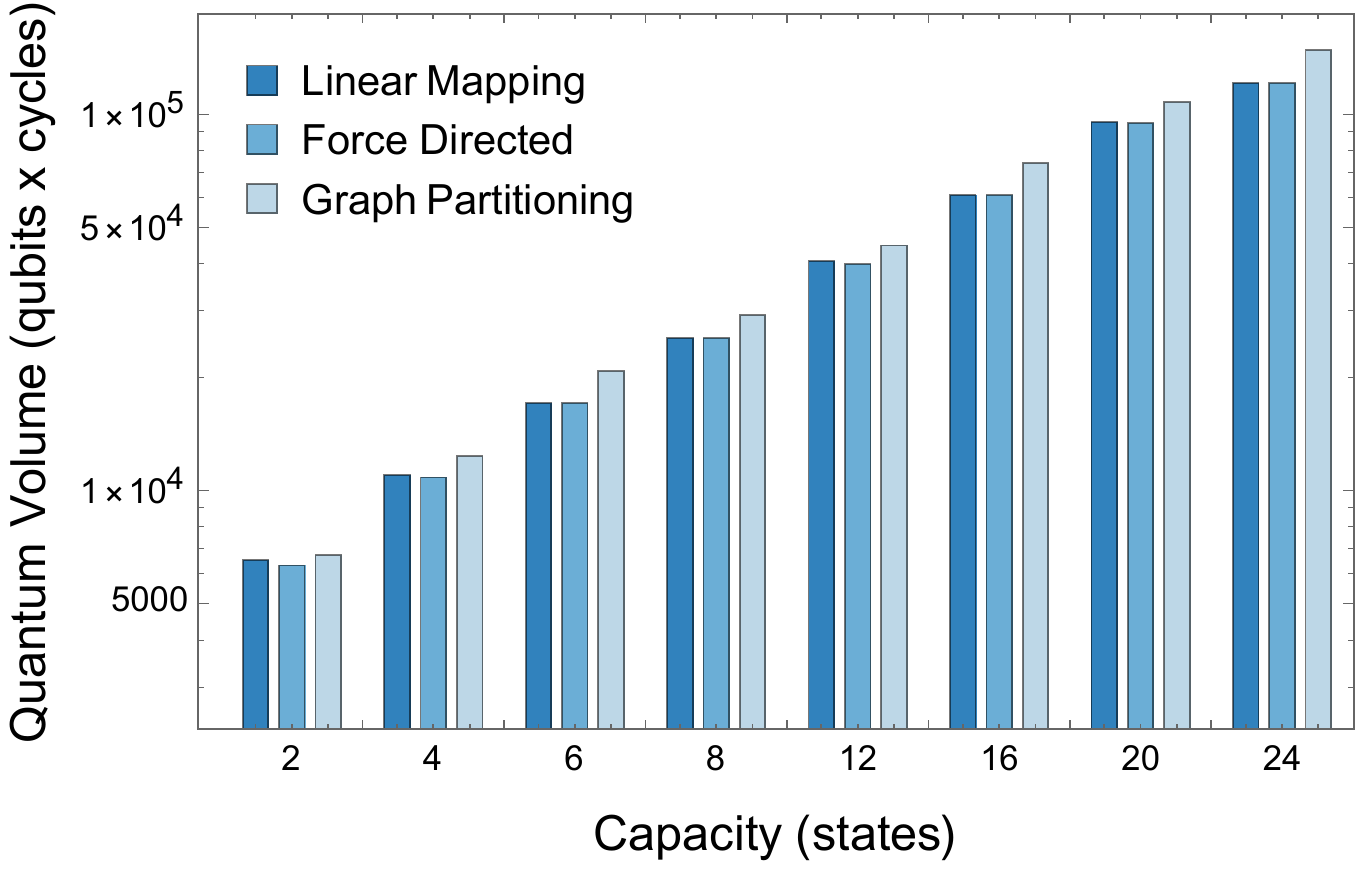}
    \caption{}
    \label{fig:l1vols}
    \end{subfigure}
    ~\hspace{1in}
    \begin{subfigure}[b]{0.33\linewidth}
    \includegraphics[width=\textwidth]{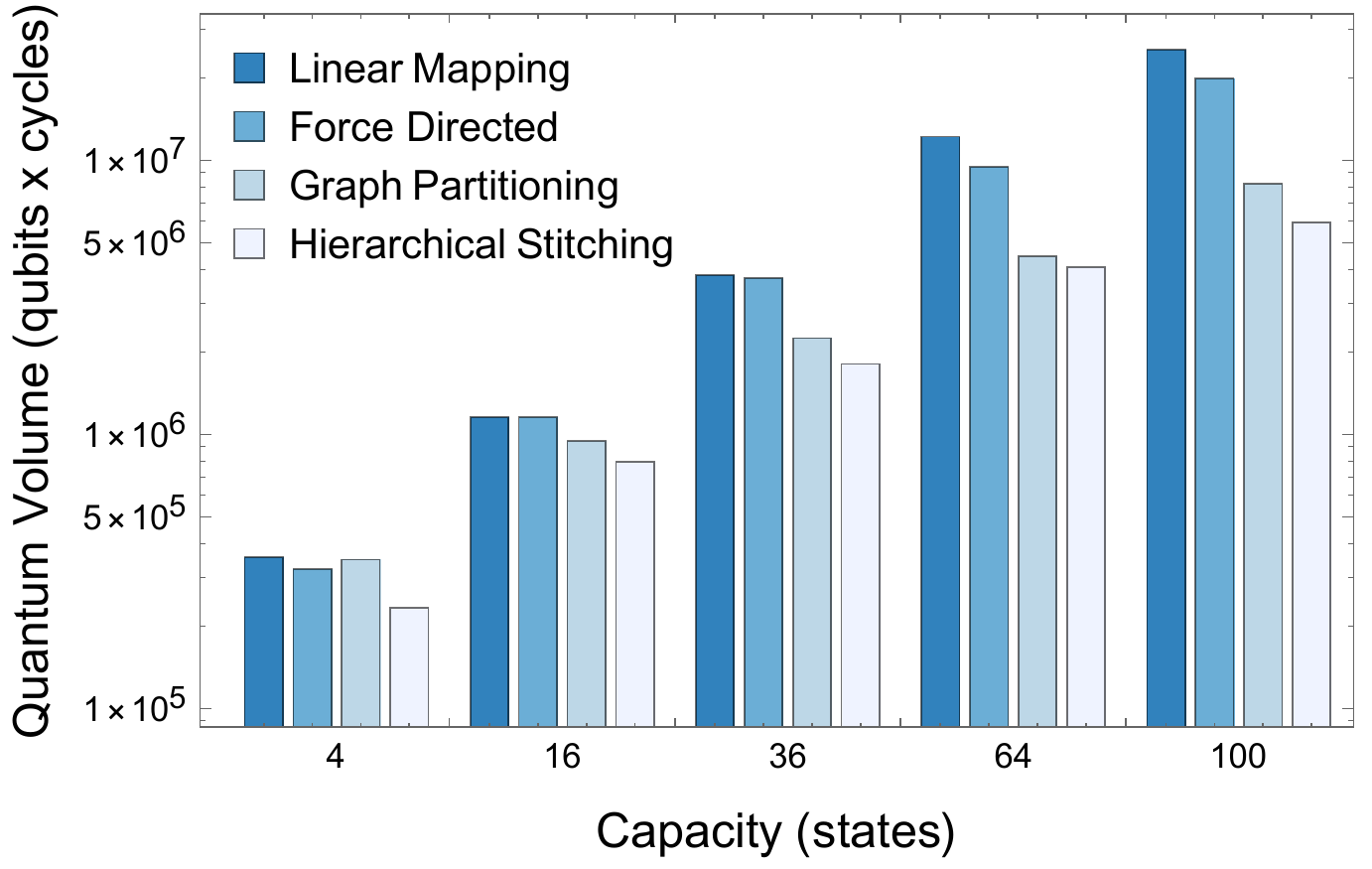}
    \caption{}
    \label{fig:l2vols}
    \end{subfigure}
    \caption{One and two level factory resource requirements. For the single level factory, we present latencies (\ref{fig:l1latencies}), areas (\ref{fig:l1areas}), and achieved space-time volumes (\ref{fig:l1vols}). For the two-level factory, the right-hand side shows latencies (\ref{fig:l2latencies}), areas (\ref{fig:l2areas}), and volumes (\ref{fig:l2vols}). All three optimizations are effective for reducing the overhead of single level factories. For two level factories, each procedure trades off space and time separately, resulting in the lowest achievable volume by that procedure. Hierarchical stitching is able to reduce space-time volume by 5.64x.}
    \label{fig:results}
\end{figure*}

\sisetup{round-precision=3, round-mode=figures, scientific-notation=true}
\begin{table*}[t]
\scriptsize
\centering
\begin{tabular}{lcccccccccc}
\hline\hline
&  \multicolumn{5}{c}{Level 1} &   \multicolumn{5}{c}{Level 2} \\
Procedure    &   $K=2$ &   4   &   8 &   10 &   24 &   $K=4$ &   16 &   36 &   64 &   100  \\
\hline
Random & \num{11075}  & \num{18216}  & \num{54272} & \num{64000} & \num{269568} & $-$ & $-$ & $-$ & $-$ & $-$ \\
Line(NR) & \num{6525}  & \num{10980}  & \num{25344} & \num{29376} & \num{128736} & \num{368400} & \num{1190856} & \num{4193376} & \num{12516194} & \num{33444400}\\
Line(R) & \num{6525}  & \num{10980}  & \num{25344} & \num{29376} & \num{128736} & \num{354502} & \num{1151904} & \num{3800928} & \num{12155724} & \num{25282500}\\
FD & \num{6300}  & \num{10800} &  \num{25344} & \num{28800} & \num{120672} & \num{322012} & \num{1151904} & \num{3716856} & \num{9447893} & \num{19816775}\\
GP & \num{6725}  & \num{12312} &  \num{29120} & \num{33344} & \num{147744} & \num{348365} & \num{940992} & \num{2239172} & \num{4448808} & \num{8172500}\\
HS & $-$  & $-$ & $-$ &  $-$ & $-$ & \num{231700} & \num{793440} & \num{1801436} & \num{4058880} & \num{5926784}\\
Critical & \num{6279}  & \num{10725}  & \num{22737} & \num{30303} & \num{112385} & \num{182234} & \num{448188} & \num{885118} & \num{1532960} & \num{2431650}\\
\hline\hline
\end{tabular}
\caption{Quantum volumes required by factory designs optimized by: randomization (Random), linear mapping (Line) with and without qubit reuse (R, NR), force-directed (FD), graph partitioning (GP), and hierarchical stitching (HS).}
\end{table*}
\subsection{Evaluation Methodology: Simulation Environment}
To perform evaluation of our methods, we implemented each configuration of the full Bravyi-Haah distillation protocol in the Scaffold programming language, and compiled this to gate-level instructions (e.g. Fig.~\ref{fig:pseudocode}). These instructions are fed into a cycle-accurate network simulator \cite{javadi2017optimized} that accurately executes the scheduling and routing of braids on a 2-dimensional surface-code qubit mesh. We extended both of these tools to support a multi-target CNOT gate. The simulator first schedules braids in parallel where the interaction graph allows. If braids intersect on the machine, the simulator inserts a stall to allow one braid to complete before the other. To perform scheduling, the simulator treats any data hazard (i.e. the presence of the same qubit in consecutive instructions) as a true dependency. This eliminates gate-level optimizations from being automatically performed, but it simultaneously allows for the introduction of “barrier” type gates. These are implemented by inserting a single gate involving all qubits of the machine (specifically a multi-target CNOT operation controlled on an extra qubit set to the zero state, and targeting all other qubits of the machine). Gate-level optimizations involving the commutativity relations of specific operations are performed by hand, independent of scheduling.

\subsection{Single-Level Factory Evaluation}
 We notice first that the linear mapping procedure \cite{fowler2013surface} performs well, even as the capacity of the factory increases. In Fig. \ref{fig:l1gap}, we see that the linear mapping technique actually is able to approach the theoretical minimum required latency for each of these circuits. These mappings were specifically designed to optimize for these single level factories, which justifes these scaling properties.

Our proposed force-directed mapping approach described in section \ref{subsec:forcedirected} is able to improve slightly from the linear mapping technique in most cases. This is due to the strong correlation of the metrics that the approach optimizes, to the realized circuit latency.

 Graph partitioning techniques described in \ref{subsec:graphpartitioning} underperform the linear mapping and force directed procedures, although they are still competent with respect to the theoretical minimum resource requirements. Because of the simplicity of the circuit and the targeted optimizations that were perform specifically for these small circuits, the advantage from the global nature of the graph partitioning method is significantly diminished. 

The realistic circuit latency as executed in simulation, required circuit area, and corresponding quantum volume for single level magic state distillation factories are shown in Fig. \ref{fig:l1latencies}, \ref{fig:l1areas}, and \ref{fig:l1vols}. The best performing approach for each of the single level factories closely approximates the theoretical minimum latency and space-time volume required by these circuits. This is leveraged by our iterative procedure and used to ultimately achieve the most efficient circuit expression.

\subsection{Multi-Level Factory Evaluation}
\subsubsection{Effects of Qubit Reuse Protocols}
As anticipated, by electing to reuse qubits for later rounds in the distillation circuits, the overall circuit consumes less area at the cost of higher latency. Qubit reuse, for both the linear mapping and graph partitioning optimization methods, results in lower space-time volume.

The force directed procedure actually achieves a lower volume when qubits are not reused. This is due to two factors introduced by qubit reuse. First, the average degree of the interaction graph has increased due to the introduction of false dependencies. This restricts the optimization procedure from being able to minimize the heuristics cleanly, as each qubit is more tightly connected to others, reducing the degrees of freedom in the graph. Second, there is more area in the graph, which widens the search space available for the procedure. With more possible configurations, the algorithm is more likely to find more optimized mappings.

\subsubsection{Optimization Procedure Comparison}
 Fig. \ref{fig:results} shows the minimized space-time volumes achieved by each optimization procedure. While the linear mapping and force-directed procedures were able to nearly optimally map single level factories, the performance deteriorates significantly when moving to multi-level factories. In these factories, Hierarchical Stitching is able to outperform the other optimization strategies, as it synthesizes the best performing components of each.

We also considered both qubit reuse and non-reuse policies. The optimal combinations vary slightly for each procedure: the linear mapping and graph partitioning strategies always perform best with qubit reuse, while the force directed procedure performs best with qubit reuse for capacity 4 and 16 two level factories, and without qubit reuse for capacity 36 and beyond. This is due to the additional degrees of freedom that avoiding qubit reuse injects, as discussed above. The final results plots show these configurations. 

In moving to multi-level factory circuits, even though there is significant modularity and symmetry in the factory, the introduction of the output state permutation from one level to the input states of the next introduces severe latency overheads. Without taking this into consideration, the linear mapping procedures suffer from large latency expansions in attempting to execute multi-level circuits, with the effect compounding as the size (output capacity) of the factory increases. Fig.~\ref{fig:l2vols} shows that the force-directed approach is able to improve to a maximum reduction of $\sim$1.27x from these linear mappings, but is constrained by how poorly these mappings originally perform.

The graph partitioning technique is able to simultaneously optimize for the entirety of the multi-level circuit, including the inter-round communication steps. With all of this information, the technique is able to minimize interaction graph edge crossings and edge lengths, which results in a more efficient expression of the circuits overall for larger two level circuits. Smaller two level circuits are still dominated by the intra-round execution overheads, which are able to be effectively minimized by linear mapping and force directed techniques. Once multi-level factories become large enough (occurring in Fig.~\ref{fig:l2vols} at capacity 16), the inter-round effects begin to dominate. This is the point when graph partitioning techniques are able to outperform other methods.

The proposed hierarchical stitching technique is able to leverage the strengths of the force directed and graph partitioning methods to more effectively reduce resource consumption by mapping each round to near optimality and utilizing the same force-directed technique combined with the introduction of intermediate destinations to mitigate the overheads incurred by inter-round communication. Within all explored multi-level factory circuits, these optimizations further reduced resource consumption. In the largest case, a capacity 100 two level factory shows a 5.64x reduction in overall consumed quantum volume when moving from the linear mapping approach without reusing qubits, to hierarchical stitching.


%


\section{Future Work}\label{sec:future}
There are a number of immediate extensions to this work:

\begin{itemize}
    \item \emph{System-Level Performance}. We are studying the effect of higher-level factory optimizations on application performance. This includes analysis of resource distribution, comparison of factory system layout topologies, as well as architectures with prepared state buffers. The interaction with the Hierarchical Stitching procedure is currently being analyzed.
    \item \emph{Stitching Generalization}. Our proposed hierarchical stitching procedure can be applied to other hierarchical circuits, and to arbitrary circuits coupled with procedures that detect hierarchical sub-circuits. For example, we may extract sets of (planar) sub-divisions from the interaction graph and map each sub-division onto the 2-D surface, and perform permutations ($\mathtt{swap}$ gates) that patches the set of mappings together.
    \item \emph{Teleportation vs. Lattice Surgery vs. Braiding}. Along the lines of \cite{javadi2017optimized,campbell}, we plan to explore the impacts of changing the surface code interaction style. Our proposed optimizations may likely change the trade off thresholds presented in \cite{javadi2017optimized}.
    \item \emph{Loss Compensation}. Typically in distillation protocols, when magic states are marked as defective they would be discarded and cause module failure. Future work would include implementing protocols that \emph{compensates} the loss of those defective magic states by having back-up maintenance modules that feed high-fidelity states to ensure the completion of the distillation round.
    \item \emph{Area Expansion}. It is possible to expand the utilized area for these distillation circuits and reduce the execution latency. Our force directed procedures work well with additional area, so this may reduce overall consumed space-time volume.
\end{itemize}
\section{Conclusion}\label{sec:conclusion}
Error correction is the largest performance bottleneck of long-term quantum computers, and magic-state distillation is the most expensive component. Known optimized scheduling and mapping techniques for state distillation circuits tend to work well for small, single level factories, but quickly incur large overheads for larger factories. We have proposed a technique that synthesizes mapping and scheduling optimizations to take advantage of the unique efficiencies of each, which allows for a significant 5.64x reduction in the realistic space-time volume required to implement multi-level magic state distillation factories. Global optimizations like graph partitioning and force-directed annealing work well, but leveraging structure of the block code circuitry combined with the specific strengths of both graph partitioning and force-directed annealing allows for the most improvement, resulting in large factors of resource reduction overall.

\section*{Acknowledgment}
This work is funded in part by EPiQC, an NSF Expedition in Computing, under grants CCF-1730449 and CCF-1730082, Los Alamos National Laboratory and the U.S. Department of Defense under subcontract 431682, by NSF PHY grant 1660686, and by a research gift from Intel Corporation.


\bibliographystyle{IEEEtran.bst}
\bibliography{references}

\end{document}